\documentstyle[12pt]{article}
\begin{document}

\begin{flushright}
{\bf hep-ph/9907455} \\
{\bf LMU-99-10} \\
July 1999
\end{flushright}

\vspace{0.2cm}

\begin{center}
{\large\bf $CP$ Violation 
in $B_d \rightarrow D^+D^-$, $D^{*+}D^-$, $D^+D^{*-}$
and $D^{*+}D^{*-}$ Decays}
\end{center}

\vspace{.5cm}
\begin{center}
{\bf Zhi-zhong Xing} \footnote{
E-mail: Xing$@$hep.physik.uni-muenchen.de }\\
{\sl Sektion Physik, Universit${\sl\ddot a}$t M${\sl\ddot u}$nchen, 
Theresienstrasse 37A, 80333 M${\sl\ddot u}$nchen, Germany}
\end{center}

\vspace{2.cm}

\begin{abstract}
$CP$ asymmetries in 
$B_d \rightarrow D^+D^-$, $D^{*+}D^-$,
$D^+D^{*-}$ and $D^{*+}D^{*-}$ decays are investigated
with the help of the factorization approximation and isospin relations.
We find that the direct $CP$ violation is governed only by
the short-distance penguin mechanism, while the indirect 
$CP$ asymmetries in $B_d \rightarrow D^{\pm}D^{*\mp}$ 
transitions may be modified due to the final-state 
rescattering effect. An updated numerical analysis shows that
the direct $CP$ asymmetry in $B^0_d$ vs $\bar{B}^0_d \rightarrow D^+D^-$ decays
can be as large as $3\%$. The $CP$-even and $CP$-odd
contributions to the indirect $CP$ asymmetry in 
$B^0_d$ vs $\bar{B}^0_d \rightarrow D^{*+}D^{*-}$
decays are found to have the rates $89\%$ and $11\%$, respectively. 
Some comments on the possibilities
to determine the weak phase $\beta$ and to test the
factorization hypothesis are also given.
\end{abstract}

\vspace{4cm}
\begin{center}
{PACS number(s): ~ 13.25.+m, 11.30.Er, 12.15.Ff, 14.40.Jz}
\end{center}

\newpage
\section{Introduction}

A direct measurement of the $CP$-violating parameter
$\sin 2\beta$ in $B^0_d$ vs $\bar{B}^0_d \rightarrow
J/\psi K_S$ decays, where 
$\beta \equiv \arg [-(V^*_{tb}V_{td})/(V^*_{cb}V_{cd})]$ 
is known as 
an inner angle of the Cabibbo-Kobayashi-Maskawa (CKM) unitarity triangle
\begin{equation}
V^*_{ub}V_{ud} ~ + ~ V^*_{cb}V_{cd} ~ + ~ V^*_{tb}V_{td} 
~ = ~ 0 ~ ,
\end{equation}
has recently been reported by the CDF Collaboration \cite{CDF}.
The preliminary result 
$\sin2\beta = 0.79^{+0.41}_{-0.44} ~ ({\rm stat + syst})$
is consistent very well with the standard-model prediction
for $\sin 2\beta$, obtained indirectly from a global analysis of
current data on $|V_{ub}/V_{cb}|$, $B^0_d$-$\bar{B}^0_d$ mixing, 
and $CP$ violation in $K^0$-$\bar{K}^0$ mixing \cite{Stocchi}.
If the CDF measurement 
is confirmed, $CP$ violation of the magnitude $\sin 2\beta$
should also be seen in the decay modes
$B^0_d$ vs $\bar{B}^0_d \rightarrow D^+D^-$, $D^{*+}D^-$,
$D^+D^{*-}$ and $D^{*+}D^{*-}$, whose branching ratios are
all anticipated to be of $O(10^{-4})$. Indeed the channel
$B^0_d \rightarrow D^{*+}D^{*-}$ has been observed 
by the CLEO Collaboration \cite{CLEO}, and the measured 
branching ratio 
${\cal B} (D^{*+}D^{*-}) = 
[ 6.2^{+4.0}_{-2.9} ~ ({\rm stat}) \pm 1.0 ~ ({\rm syst})]
\times 10^{-4}$
is in agreement with the standard-model
expectation. Further measurements of neutral and charged $B$
decays into $D^{(*)} \bar{D}^{(*)}$ states will soon be
available in the first-round experiments of KEK and SLAC
$B$-meson factories as well as at other high-luminosity hadron
machines (see, e.g., Ref. \cite{BB} for a review with
extensive references).

\vspace{0.4cm}

In the literature some special attention has been paid to 
$B \rightarrow D^{(*)}\bar{D}^{(*)}$ transitions and $CP$
violation. For example, the $CP$ properties of 
$B_d \rightarrow D^{(*)+} D^{(*)-}$ decays were analyzed
in the heavy quark limit in Ref. \cite{Aleksan}; the
isospin relations and penguin effects in $B\rightarrow
D^{(*)}\bar{D}^{(*)}$ decays were explored in Ref. \cite{Xing97};
the possibility of extracting the weak phase $\beta$ and
testing the factorization hypothesis in $B^0_d$ vs $\bar{B}^0_d$
decays into the non-$CP$ eigenstates $D^{\pm}D^{*\mp}$ were
investigated in Ref. \cite{Xing98}; and the angular analysis
of $B_d \rightarrow D^{*+}D^{*-}$ decays to determine 
$CP$-even and $CP$-odd amplitudes were presented in 
Ref. \cite{Dunietz}. In addition to those works, 
numerical estimates of branching ratios and $CP$ asymmetries
in $B\rightarrow D^{(*)}\bar{D}^{(*)}$ decays have been
given in Ref. \cite{Number}, in which neither electroweak penguin
contributions nor final-state rescattering effects were taken
into account.

\vspace{0.4cm}

The present paper, different in several aspects from those previous studies,
aims at analyzing final-state rescattering effects on
direct and indirect $CP$ asymmetries in $B\rightarrow
D^{(*)}\bar{D}^{(*)}$ decays.
We calculate the $I=1$ and $I=0$ isospin amplitudes of
these processes by using the factorization approximation and
the effective weak Hamiltonian, and account for 
long-distance interactions at the hadron level
by introducing elastic rescattering
phases for two isospin channels of the final-state mesons.
In this approach we find that direct $CP$ asymmetries in
both charged and neutral $B$ decay modes are governed only
by the short-distance penguin mechanism, but indirect $CP$
asymmetries in $B_d \rightarrow D^{\pm}D^{*\mp}$ transitions
may be modified due to the final-state rescattering effect.
An updated numerical analysis of direct $CP$ violation in
$B\rightarrow D\bar{D}$, $D^*\bar{D}$, $D\bar{D}^*$ and
$D^*\bar{D}^*$ decays is made without neglect of the 
electroweak penguin effects. We obtain the asymmetry as large
as $3\%$ in $B^+_u \rightarrow D^+\bar{D}^0$ vs
$B^-_u\rightarrow D^-D^0$ or $B^0_d$ vs $\bar{B}^0_d
\rightarrow D^+D^-$ decays. 
In the absence of angular
analysis we find that the indirect $CP$ asymmetry in
$B_d\rightarrow D^{*+}D^{*-}$ decays is diluted by a factor
0.89, i.e., $11\%$ of the asymmetry 
arising from the $P$-wave ($CP$-odd) contribution.
We also give some comments on the possibilities to
determine the weak phase $\beta$ and to test the factorization
hypothesis in the presence of final-state interactions.

\section{Isospin amplitudes}

The effective weak Hamiltonian responsible for 
$B\rightarrow D^{(*)} \bar{D}^{(*)}$ decays can explicitly
be written as \cite{Buras}
\begin{equation}
{\cal H}_{\rm eff} \; =\; \frac{G_{\rm F}}{\sqrt{2}}
\sum_q \left [ V_{qb}V^*_{qd} \left ( \sum^2_{i=1}
c_i Q^q_i + \sum^{10}_{i=3} c_i Q_i \right ) \right ] 
~ + ~ {\rm h.c.} \; ,
\end{equation}
where $V_{qb}$ and $V_{qd}$ (for $q=u, c$) are the CKM matrix
elements, $c_i$ (for $i=1, \cdot\cdot\cdot, 10$)
are the Wilson coefficients, and
\begin{eqnarray}
Q^q_1 & = & (\bar{d}_\alpha q_\beta )^{~}_{\rm V-A}
(\bar{q}_\beta b_\alpha )^{~}_{\rm V-A} \; ,
\nonumber \\
Q^q_2 & = & (\bar{d}q )^{~}_{\rm V-A} 
(\bar{q}b )^{~}_{\rm V-A} \; , 
\nonumber \\
Q_3 & = & (\bar{d}b )^{~}_{\rm V-A} 
(\bar{c}c )^{~}_{\rm V-A} \; ,
\nonumber \\
Q_4 & = & (\bar{d}_\alpha b_\beta )^{~}_{\rm V-A}
(\bar{c}_\beta c_\alpha )^{~}_{\rm V-A} \; ,
\nonumber \\
Q_5 & = & (\bar{d}b )^{~}_{\rm V-A} 
(\bar{c}c )^{~}_{\rm V+A} \; ,
\nonumber \\
Q_6 & = & (\bar{d}_\alpha b_\beta )^{~}_{\rm V-A}
(\bar{c}_\beta c_\alpha )^{~}_{\rm V+A} \; ,
\end{eqnarray}
as well as $Q_7 = Q_5$, $Q_8 = Q_6$, $Q_9 = Q_3$ and $Q_{10} = Q_4$. 
Here $Q_3, \cdot\cdot\cdot ,Q_6$ denote the QCD-induced penguin
operators, and $Q_7, \cdot\cdot\cdot ,Q_{10}$ stand for the electroweak
penguin operators. 

\vspace{0.4cm}

It is clear that the 
$\Delta B =+1$ and $\Delta B =-1$ parts of 
${\cal H}_{\rm eff}$ have the isospin 
structures $|1/2, +1/2\rangle$ and $|1/2, -1/2\rangle$,
respectively. They govern the transitions
$B^+_u\rightarrow D^{(*)+}\bar{D}^{(*)0}$, $B^0_d\rightarrow
D^{(*)+}D^{(*)-}$, $B^0_d\rightarrow D^{(*)0}
\bar{D}^{(*)0}$ and their $CP$-conjugate processes. The
final state of each decay mode 
can be in either $I=1$ or $I=0$ isospin configuration.
For simplicity we denote the amplitudes of six relevant 
transitions by use of the electric charges of their final-
and initial-state mesons,
i.e., $A^{+0}$, $A^{+-}$, $A^{00}$ (for $B^+_u$ and $B^0_d$ decays) 
and $\bar{A}^{-0}$, $\bar{A}^{+-}$,
$\bar{A}^{00}$ (for $B^-_u$ and $\bar{B}^0_d$ decays).
These amplitudes can be expressed in terms of the $I=1$ and $I=0$
isospin amplitudes, which include both weak and strong phases.
For example \cite{Xing97}
\footnote{As for $B\rightarrow D^*\bar{D}^*$ decays, the 
isospin relations hold separately for the transition amplitudes
with helicity $\lambda = -1$, 0 or $+1$.},
\begin{eqnarray}
A^{+0} & = & A_1 \; , \nonumber \\
A^{+-} & = & \frac{1}{2} (A_1 + A_0 ) \; , \nonumber \\
A^{00} & = & \frac{1}{2} (A_1 - A_0 ) \; ;
\end{eqnarray}
and the relations between 
($\bar{A}^{-0}$, $\bar{A}^{+-}$, $\bar{A}^{00}$) and
($\bar{A}_1$, $\bar{A}_0$) hold in the same form.
In the complex plane two sets of isospin relations form two 
triangles: $A^{+0} = A^{+-} + A^{00}$ and 
$\bar{A}^{-0} = \bar{A}^{+-} + \bar{A}^{00}$.

\vspace{0.4cm}

To calculate the magnitudes of $I=1$ and $I=0$ isospin amplitudes, 
we make use of the effective Hamiltonian ${\cal H}_{\rm eff}$ and the
factorization approximation. We neglect the contributions of
the annihilation-type channels, which are expected
to have significant form-factor suppression \cite{Xing96}. 
It should be noted
that in this approach the Wilson coefficients and the 
relevant hadronic matrix elements of four-quark operators 
need be evaluated in the same renormalization scheme and
at the same energy scale. Following the procedure
described in Ref. \cite{Fleischer} one can obtain the scale- and
renormalization-scheme--independent transition amplitudes
consisting of the CKM factors, the
effective Wilson coefficients, the penguin loop-integral functions
and the factorized hadronic matrix elements.
Under isospin symmetry, we are only left with two different
hadronic matrix elements:
\begin{eqnarray}
Z & = & \langle D^{(*)+}|(\bar{c}d)^{~}_{\rm V-A}|0\rangle
\langle D^{(*)-}|(\bar{b}c)^{~}_{\rm V-A}|B^0_d\rangle
\nonumber \\
& = & \langle D^{(*)+}|(\bar{c}d)^{~}_{\rm V-A}|0\rangle
\langle \bar{D}^{(*)0}|(\bar{b}c)^{~}_{\rm V-A}|B^+_u\rangle
\; , \nonumber \\
\bar{Z} & = & \langle D^{(*)-}|(\bar{d}c)^{~}_{\rm V-A}|0\rangle
\langle D^{(*)+}|(\bar{c}b)^{~}_{\rm V-A}|\bar{B}^0_d\rangle
\nonumber \\
& = & \langle D^{(*)-}|(\bar{d}c)^{~}_{\rm V-A}
|0\rangle \langle D^{(*)0}|(\bar{c}b)^{~}_{\rm V-A}|B^-_u\rangle \; . 
\end{eqnarray}
Note that $|\bar{Z}|=|Z|$ holds for the final states with two
pseudoscalar mesons or those with one pseudoscalar and one vector
mesons. Only for the final states
with two vector mesons $|\bar{Z}|$ and $|Z|$ are different,
as the $P$-wave contributions to $Z$ and $\bar{Z}$
have the opposite signs (see section 4 for the detail). 
Furthermore, we account for final-state
interactions at the hadron level by introducing the elastic
rescattering phases $\delta_1$ and $\delta_0$ for 
$I=1$ and $I=0$ isospin channels (a similar treatment
can be found, e.g., in Refs. \cite{Deshpande,Neubert}). 
We then arrive
at the factorized isospin amplitudes as follows:
\begin{eqnarray}
A_1 & = & \frac{G_{\rm F}}{\sqrt{2}}
\left (V_{ud}V_{ub}^* S_u + V_{cd}V_{cb}^* S_c \right )
Z e^{{\rm i}\delta_1} \; , \nonumber \\
A_0 & = & \frac{G_{\rm F}}{\sqrt{2}}
\left (V_{ud}V_{ub}^* S_u + V_{cd}V_{cb}^* S_c \right )
Z e^{{\rm i}\delta_0} \; ,
\end{eqnarray}
in which $S_u$ and $S_c$ are composed of the effective
Wilson coefficients and the penguin loop-integral functions
(see section 3). The expressions of $\bar{A}_1$ and $\bar{A}_0$ can
be obtained respectively from those of $A_1$ and $A_0$ in
Eq. (6) through the replacements $Z\Longrightarrow \bar{Z}$ and
$V_{qd}V^*_{qb} \Longrightarrow V^*_{qd}V_{qb}$ (for $q=u$ and $c$).
Note that all parameters in the isospin amplitudes,
except the CKM factors, are dependent upon the specific 
final states of $B$ decays.

\vspace{0.4cm}

One can see that $|A_0| = |A_1|$ and $|\bar{A}_0|=|\bar{A}_1|$
hold in the context of our simple factorization scheme.
This implies that the $B_d\rightarrow D^{(*)0} \bar{D}^{(*)0}$
transitions would be forbidden, if there were no final-state
rescattering effects (i.e., if $\delta_0 = \delta_1$). 
Substituting Eq. (6) into Eq. (4), one obtains
\begin{eqnarray}
A^{+-} & = & ~ \frac{G_{\rm F}}{\sqrt{2}}
\left (V_{ud}V^*_{ub} S_u + V_{cd}V^*_{cb} S_c
\right ) Z \cos\frac{\delta_1 -\delta_0}{2} 
e^{{\rm i} (\delta_1 + \delta_0)/2} \; , 
\nonumber \\
A^{00} & = & {\rm i} \frac{G_{\rm F}}{\sqrt{2}}
\left (V_{ud}V^*_{ub} S_u + V_{cd}V^*_{cb} S_c
\right ) Z \sin\frac{\delta_1 -\delta_0}{2} 
e^{{\rm i} (\delta_1 + \delta_0)/2} \; .
\end{eqnarray}
Similarly $\bar{A}^{+-}$ and $\bar{A}^{00}$ can be read off from
$A^{+-}$ and $A^{00}$ through the replacements
$Z\Longrightarrow \bar{Z}$ and
$V_{qd}V^*_{qb} \Longrightarrow V^*_{qd}V_{qb}$ (for $q=u$ and $c$).
It is easy to find 
\begin{eqnarray}
|A^{+-}|^2 + |A^{00}|^2 & = & |A^{+0}|^2 \; ,
\nonumber \\
|\bar{A}^{+-}|^2 + |\bar{A}^{00}|^2 & = & |\bar{A}^{-0}|^2 \; ;
\end{eqnarray}
i.e., the two isospin triangles are right-angled triangles.
Whether the relations in Eq. (8) are practically valid or not
can be checked, once the
experimental data on branching ratios of $B\rightarrow
D^{(*)}\bar{D}^{(*)}$ decays are available. 

\vspace{0.4cm}

If $|A^{+-}| = |A^{+0}|$ held, $|A^{00}| =0$ would result
within the factorization approach described above. Namely,
observation of the (approximate) equality between the decay
rates of $B^0_d\rightarrow D^{(*)+}D^{(*)-}$ and
$B^+_u\rightarrow D^{(*)+}\bar{D}^{(*)0}$ would imply that
the decay modes $B^0_d\rightarrow D^{(*)0}\bar{D}^{(*)0}$
were forbidden or strongly suppressed. This conclusion is 
in general not true, however. 
Without any special assumption or approximation,
we denote $A_0/A_1 = z e^{{\rm i}\theta}$,
$|A^{00}/A^{+0}|^2 = R$ and obtain consequences
of the equality $|A^{+-}| = |A^{+0}|$ as follows:
\begin{eqnarray}
z & = & \sqrt{3 + \cos^2\theta} ~ - ~ \cos\theta \; , \nonumber \\
R & = & 1 + \cos^2 \theta ~ - ~ \cos\theta \sqrt{3 + \cos^2\theta} \; .
\end{eqnarray}
The behaviors of $z$ and $R$ changing with $\theta$ is illustrated
in Fig. 1. It is clear that in general
$|A^{00}| =0$ (i.e., $R=0$) 
is not necessary to hold even if $|A^{+-}| =|A^{+0}|$ holds.
Therefore the detection of $B_d \rightarrow D^{(*)0}\bar{D}^{(*)0}$
transitions is very useful in experiments, in order to demonstrate
whether final-state rescattering effects are significant and to test
whether the factorization approximation works well.
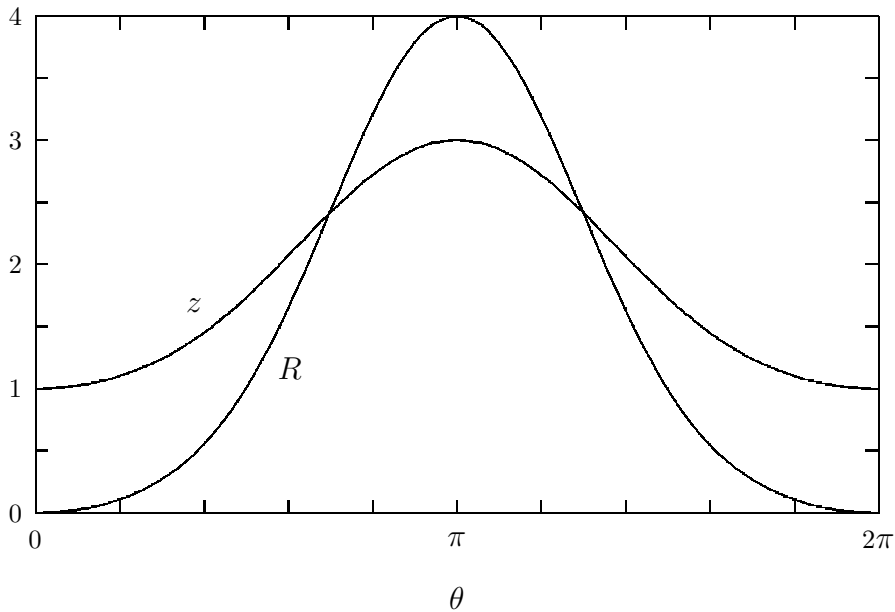
\begin{figure}
\setlength{\unitlength}{0.240900pt}
\ifx\plotpoint\undefined\newsavebox{\plotpoint}\fi
\sbox{\plotpoint}{\rule[-0.200pt]{0.400pt}{0.400pt}}%
\begin{picture}(1424,944)(-250,0)
\font\gnuplot=cmr10 at 10pt
\gnuplot
\sbox{\plotpoint}{\rule[-0.200pt]{0.400pt}{0.400pt}}%
\put(100.0,123.0){\rule[-0.200pt]{4.818pt}{0.400pt}}
\put(80,123){\makebox(0,0)[r]{0}}
\put(1404.0,123.0){\rule[-0.200pt]{4.818pt}{0.400pt}}
\put(100.0,221.0){\rule[-0.200pt]{4.818pt}{0.400pt}}
\put(1404.0,221.0){\rule[-0.200pt]{4.818pt}{0.400pt}}
\put(100.0,318.0){\rule[-0.200pt]{4.818pt}{0.400pt}}
\put(80,318){\makebox(0,0)[r]{1}}
\put(1404.0,318.0){\rule[-0.200pt]{4.818pt}{0.400pt}}
\put(100.0,416.0){\rule[-0.200pt]{4.818pt}{0.400pt}}
\put(1404.0,416.0){\rule[-0.200pt]{4.818pt}{0.400pt}}
\put(100.0,514.0){\rule[-0.200pt]{4.818pt}{0.400pt}}
\put(80,514){\makebox(0,0)[r]{2}}
\put(1404.0,514.0){\rule[-0.200pt]{4.818pt}{0.400pt}}
\put(100.0,611.0){\rule[-0.200pt]{4.818pt}{0.400pt}}
\put(1404.0,611.0){\rule[-0.200pt]{4.818pt}{0.400pt}}
\put(100.0,709.0){\rule[-0.200pt]{4.818pt}{0.400pt}}
\put(80,709){\makebox(0,0)[r]{3}}
\put(1404.0,709.0){\rule[-0.200pt]{4.818pt}{0.400pt}}
\put(100.0,806.0){\rule[-0.200pt]{4.818pt}{0.400pt}}
\put(1404.0,806.0){\rule[-0.200pt]{4.818pt}{0.400pt}}
\put(100.0,904.0){\rule[-0.200pt]{4.818pt}{0.400pt}}
\put(80,904){\makebox(0,0)[r]{4}}
\put(1404.0,904.0){\rule[-0.200pt]{4.818pt}{0.400pt}}
\put(100.0,123.0){\rule[-0.200pt]{0.400pt}{4.818pt}}
\put(100,82){\makebox(0,0){0}}
\put(100.0,884.0){\rule[-0.200pt]{0.400pt}{4.818pt}}
\put(232.0,123.0){\rule[-0.200pt]{0.400pt}{4.818pt}}
\put(232.0,884.0){\rule[-0.200pt]{0.400pt}{4.818pt}}
\put(365.0,123.0){\rule[-0.200pt]{0.400pt}{4.818pt}}
\put(365.0,884.0){\rule[-0.200pt]{0.400pt}{4.818pt}}
\put(497.0,123.0){\rule[-0.200pt]{0.400pt}{4.818pt}}
\put(497.0,884.0){\rule[-0.200pt]{0.400pt}{4.818pt}}
\put(630.0,123.0){\rule[-0.200pt]{0.400pt}{4.818pt}}
\put(630.0,884.0){\rule[-0.200pt]{0.400pt}{4.818pt}}
\put(762.0,123.0){\rule[-0.200pt]{0.400pt}{4.818pt}}
\put(762,82){\makebox(0,0){$\pi$}}
\put(762.0,884.0){\rule[-0.200pt]{0.400pt}{4.818pt}}
\put(894.0,123.0){\rule[-0.200pt]{0.400pt}{4.818pt}}
\put(894.0,884.0){\rule[-0.200pt]{0.400pt}{4.818pt}}
\put(1027.0,123.0){\rule[-0.200pt]{0.400pt}{4.818pt}}
\put(1027.0,884.0){\rule[-0.200pt]{0.400pt}{4.818pt}}
\put(1159.0,123.0){\rule[-0.200pt]{0.400pt}{4.818pt}}
\put(1159.0,884.0){\rule[-0.200pt]{0.400pt}{4.818pt}}
\put(1292.0,123.0){\rule[-0.200pt]{0.400pt}{4.818pt}}
\put(1292.0,884.0){\rule[-0.200pt]{0.400pt}{4.818pt}}
\put(1424.0,123.0){\rule[-0.200pt]{0.400pt}{4.818pt}}
\put(1424,82){\makebox(0,0){2$\pi$}}
\put(1424.0,884.0){\rule[-0.200pt]{0.400pt}{4.818pt}}
\put(100.0,123.0){\rule[-0.200pt]{318.952pt}{0.400pt}}
\put(1424.0,123.0){\rule[-0.200pt]{0.400pt}{188.143pt}}
\put(100.0,904.0){\rule[-0.200pt]{318.952pt}{0.400pt}}
\put(762,-10){\makebox(0,0){$\theta$}}
\put(350,450){\makebox(0,0){$z$}}
\put(500,350){\makebox(0,0){$R$}}
\put(100.0,123.0){\rule[-0.200pt]{0.400pt}{188.143pt}}
\put(100,318){\usebox{\plotpoint}}
\put(113,317.67){\rule{3.132pt}{0.400pt}}
\multiput(113.00,317.17)(6.500,1.000){2}{\rule{1.566pt}{0.400pt}}
\put(126,318.67){\rule{3.373pt}{0.400pt}}
\multiput(126.00,318.17)(7.000,1.000){2}{\rule{1.686pt}{0.400pt}}
\put(140,319.67){\rule{3.132pt}{0.400pt}}
\multiput(140.00,319.17)(6.500,1.000){2}{\rule{1.566pt}{0.400pt}}
\put(153,321.17){\rule{2.700pt}{0.400pt}}
\multiput(153.00,320.17)(7.396,2.000){2}{\rule{1.350pt}{0.400pt}}
\put(166,323.17){\rule{2.700pt}{0.400pt}}
\multiput(166.00,322.17)(7.396,2.000){2}{\rule{1.350pt}{0.400pt}}
\multiput(179.00,325.61)(2.918,0.447){3}{\rule{1.967pt}{0.108pt}}
\multiput(179.00,324.17)(9.918,3.000){2}{\rule{0.983pt}{0.400pt}}
\multiput(193.00,328.61)(2.695,0.447){3}{\rule{1.833pt}{0.108pt}}
\multiput(193.00,327.17)(9.195,3.000){2}{\rule{0.917pt}{0.400pt}}
\multiput(206.00,331.61)(2.695,0.447){3}{\rule{1.833pt}{0.108pt}}
\multiput(206.00,330.17)(9.195,3.000){2}{\rule{0.917pt}{0.400pt}}
\multiput(219.00,334.60)(1.797,0.468){5}{\rule{1.400pt}{0.113pt}}
\multiput(219.00,333.17)(10.094,4.000){2}{\rule{0.700pt}{0.400pt}}
\multiput(232.00,338.59)(1.489,0.477){7}{\rule{1.220pt}{0.115pt}}
\multiput(232.00,337.17)(11.468,5.000){2}{\rule{0.610pt}{0.400pt}}
\multiput(246.00,343.59)(1.378,0.477){7}{\rule{1.140pt}{0.115pt}}
\multiput(246.00,342.17)(10.634,5.000){2}{\rule{0.570pt}{0.400pt}}
\multiput(259.00,348.59)(1.378,0.477){7}{\rule{1.140pt}{0.115pt}}
\multiput(259.00,347.17)(10.634,5.000){2}{\rule{0.570pt}{0.400pt}}
\multiput(272.00,353.59)(1.123,0.482){9}{\rule{0.967pt}{0.116pt}}
\multiput(272.00,352.17)(10.994,6.000){2}{\rule{0.483pt}{0.400pt}}
\multiput(285.00,359.59)(1.214,0.482){9}{\rule{1.033pt}{0.116pt}}
\multiput(285.00,358.17)(11.855,6.000){2}{\rule{0.517pt}{0.400pt}}
\multiput(299.00,365.59)(0.950,0.485){11}{\rule{0.843pt}{0.117pt}}
\multiput(299.00,364.17)(11.251,7.000){2}{\rule{0.421pt}{0.400pt}}
\multiput(312.00,372.59)(0.824,0.488){13}{\rule{0.750pt}{0.117pt}}
\multiput(312.00,371.17)(11.443,8.000){2}{\rule{0.375pt}{0.400pt}}
\multiput(325.00,380.59)(0.824,0.488){13}{\rule{0.750pt}{0.117pt}}
\multiput(325.00,379.17)(11.443,8.000){2}{\rule{0.375pt}{0.400pt}}
\multiput(338.00,388.59)(0.786,0.489){15}{\rule{0.722pt}{0.118pt}}
\multiput(338.00,387.17)(12.501,9.000){2}{\rule{0.361pt}{0.400pt}}
\multiput(352.00,397.59)(0.728,0.489){15}{\rule{0.678pt}{0.118pt}}
\multiput(352.00,396.17)(11.593,9.000){2}{\rule{0.339pt}{0.400pt}}
\multiput(365.00,406.58)(0.652,0.491){17}{\rule{0.620pt}{0.118pt}}
\multiput(365.00,405.17)(11.713,10.000){2}{\rule{0.310pt}{0.400pt}}
\multiput(378.00,416.58)(0.590,0.492){19}{\rule{0.573pt}{0.118pt}}
\multiput(378.00,415.17)(11.811,11.000){2}{\rule{0.286pt}{0.400pt}}
\multiput(391.00,427.58)(0.637,0.492){19}{\rule{0.609pt}{0.118pt}}
\multiput(391.00,426.17)(12.736,11.000){2}{\rule{0.305pt}{0.400pt}}
\multiput(405.00,438.58)(0.590,0.492){19}{\rule{0.573pt}{0.118pt}}
\multiput(405.00,437.17)(11.811,11.000){2}{\rule{0.286pt}{0.400pt}}
\multiput(418.00,449.58)(0.539,0.492){21}{\rule{0.533pt}{0.119pt}}
\multiput(418.00,448.17)(11.893,12.000){2}{\rule{0.267pt}{0.400pt}}
\multiput(431.00,461.58)(0.497,0.493){23}{\rule{0.500pt}{0.119pt}}
\multiput(431.00,460.17)(11.962,13.000){2}{\rule{0.250pt}{0.400pt}}
\multiput(444.00,474.58)(0.497,0.493){23}{\rule{0.500pt}{0.119pt}}
\multiput(444.00,473.17)(11.962,13.000){2}{\rule{0.250pt}{0.400pt}}
\multiput(457.00,487.58)(0.536,0.493){23}{\rule{0.531pt}{0.119pt}}
\multiput(457.00,486.17)(12.898,13.000){2}{\rule{0.265pt}{0.400pt}}
\multiput(471.00,500.58)(0.497,0.493){23}{\rule{0.500pt}{0.119pt}}
\multiput(471.00,499.17)(11.962,13.000){2}{\rule{0.250pt}{0.400pt}}
\multiput(484.58,513.00)(0.493,0.536){23}{\rule{0.119pt}{0.531pt}}
\multiput(483.17,513.00)(13.000,12.898){2}{\rule{0.400pt}{0.265pt}}
\multiput(497.58,527.00)(0.493,0.536){23}{\rule{0.119pt}{0.531pt}}
\multiput(496.17,527.00)(13.000,12.898){2}{\rule{0.400pt}{0.265pt}}
\multiput(510.00,541.58)(0.536,0.493){23}{\rule{0.531pt}{0.119pt}}
\multiput(510.00,540.17)(12.898,13.000){2}{\rule{0.265pt}{0.400pt}}
\multiput(524.58,554.00)(0.493,0.536){23}{\rule{0.119pt}{0.531pt}}
\multiput(523.17,554.00)(13.000,12.898){2}{\rule{0.400pt}{0.265pt}}
\multiput(537.58,568.00)(0.493,0.536){23}{\rule{0.119pt}{0.531pt}}
\multiput(536.17,568.00)(13.000,12.898){2}{\rule{0.400pt}{0.265pt}}
\multiput(550.00,582.58)(0.497,0.493){23}{\rule{0.500pt}{0.119pt}}
\multiput(550.00,581.17)(11.962,13.000){2}{\rule{0.250pt}{0.400pt}}
\multiput(563.00,595.58)(0.536,0.493){23}{\rule{0.531pt}{0.119pt}}
\multiput(563.00,594.17)(12.898,13.000){2}{\rule{0.265pt}{0.400pt}}
\multiput(577.00,608.58)(0.539,0.492){21}{\rule{0.533pt}{0.119pt}}
\multiput(577.00,607.17)(11.893,12.000){2}{\rule{0.267pt}{0.400pt}}
\multiput(590.00,620.58)(0.539,0.492){21}{\rule{0.533pt}{0.119pt}}
\multiput(590.00,619.17)(11.893,12.000){2}{\rule{0.267pt}{0.400pt}}
\multiput(603.00,632.58)(0.539,0.492){21}{\rule{0.533pt}{0.119pt}}
\multiput(603.00,631.17)(11.893,12.000){2}{\rule{0.267pt}{0.400pt}}
\multiput(616.00,644.58)(0.704,0.491){17}{\rule{0.660pt}{0.118pt}}
\multiput(616.00,643.17)(12.630,10.000){2}{\rule{0.330pt}{0.400pt}}
\multiput(630.00,654.58)(0.652,0.491){17}{\rule{0.620pt}{0.118pt}}
\multiput(630.00,653.17)(11.713,10.000){2}{\rule{0.310pt}{0.400pt}}
\multiput(643.00,664.59)(0.728,0.489){15}{\rule{0.678pt}{0.118pt}}
\multiput(643.00,663.17)(11.593,9.000){2}{\rule{0.339pt}{0.400pt}}
\multiput(656.00,673.59)(0.824,0.488){13}{\rule{0.750pt}{0.117pt}}
\multiput(656.00,672.17)(11.443,8.000){2}{\rule{0.375pt}{0.400pt}}
\multiput(669.00,681.59)(1.026,0.485){11}{\rule{0.900pt}{0.117pt}}
\multiput(669.00,680.17)(12.132,7.000){2}{\rule{0.450pt}{0.400pt}}
\multiput(683.00,688.59)(0.950,0.485){11}{\rule{0.843pt}{0.117pt}}
\multiput(683.00,687.17)(11.251,7.000){2}{\rule{0.421pt}{0.400pt}}
\multiput(696.00,695.59)(1.378,0.477){7}{\rule{1.140pt}{0.115pt}}
\multiput(696.00,694.17)(10.634,5.000){2}{\rule{0.570pt}{0.400pt}}
\multiput(709.00,700.60)(1.797,0.468){5}{\rule{1.400pt}{0.113pt}}
\multiput(709.00,699.17)(10.094,4.000){2}{\rule{0.700pt}{0.400pt}}
\put(722,704.17){\rule{2.900pt}{0.400pt}}
\multiput(722.00,703.17)(7.981,2.000){2}{\rule{1.450pt}{0.400pt}}
\put(736,706.17){\rule{2.700pt}{0.400pt}}
\multiput(736.00,705.17)(7.396,2.000){2}{\rule{1.350pt}{0.400pt}}
\put(749,707.67){\rule{3.132pt}{0.400pt}}
\multiput(749.00,707.17)(6.500,1.000){2}{\rule{1.566pt}{0.400pt}}
\put(762,707.67){\rule{3.132pt}{0.400pt}}
\multiput(762.00,708.17)(6.500,-1.000){2}{\rule{1.566pt}{0.400pt}}
\put(775,706.17){\rule{2.700pt}{0.400pt}}
\multiput(775.00,707.17)(7.396,-2.000){2}{\rule{1.350pt}{0.400pt}}
\put(788,704.17){\rule{2.900pt}{0.400pt}}
\multiput(788.00,705.17)(7.981,-2.000){2}{\rule{1.450pt}{0.400pt}}
\multiput(802.00,702.94)(1.797,-0.468){5}{\rule{1.400pt}{0.113pt}}
\multiput(802.00,703.17)(10.094,-4.000){2}{\rule{0.700pt}{0.400pt}}
\multiput(815.00,698.93)(1.378,-0.477){7}{\rule{1.140pt}{0.115pt}}
\multiput(815.00,699.17)(10.634,-5.000){2}{\rule{0.570pt}{0.400pt}}
\multiput(828.00,693.93)(0.950,-0.485){11}{\rule{0.843pt}{0.117pt}}
\multiput(828.00,694.17)(11.251,-7.000){2}{\rule{0.421pt}{0.400pt}}
\multiput(841.00,686.93)(1.026,-0.485){11}{\rule{0.900pt}{0.117pt}}
\multiput(841.00,687.17)(12.132,-7.000){2}{\rule{0.450pt}{0.400pt}}
\multiput(855.00,679.93)(0.824,-0.488){13}{\rule{0.750pt}{0.117pt}}
\multiput(855.00,680.17)(11.443,-8.000){2}{\rule{0.375pt}{0.400pt}}
\multiput(868.00,671.93)(0.728,-0.489){15}{\rule{0.678pt}{0.118pt}}
\multiput(868.00,672.17)(11.593,-9.000){2}{\rule{0.339pt}{0.400pt}}
\multiput(881.00,662.92)(0.652,-0.491){17}{\rule{0.620pt}{0.118pt}}
\multiput(881.00,663.17)(11.713,-10.000){2}{\rule{0.310pt}{0.400pt}}
\multiput(894.00,652.92)(0.704,-0.491){17}{\rule{0.660pt}{0.118pt}}
\multiput(894.00,653.17)(12.630,-10.000){2}{\rule{0.330pt}{0.400pt}}
\multiput(908.00,642.92)(0.539,-0.492){21}{\rule{0.533pt}{0.119pt}}
\multiput(908.00,643.17)(11.893,-12.000){2}{\rule{0.267pt}{0.400pt}}
\multiput(921.00,630.92)(0.539,-0.492){21}{\rule{0.533pt}{0.119pt}}
\multiput(921.00,631.17)(11.893,-12.000){2}{\rule{0.267pt}{0.400pt}}
\multiput(934.00,618.92)(0.539,-0.492){21}{\rule{0.533pt}{0.119pt}}
\multiput(934.00,619.17)(11.893,-12.000){2}{\rule{0.267pt}{0.400pt}}
\multiput(947.00,606.92)(0.536,-0.493){23}{\rule{0.531pt}{0.119pt}}
\multiput(947.00,607.17)(12.898,-13.000){2}{\rule{0.265pt}{0.400pt}}
\multiput(961.00,593.92)(0.497,-0.493){23}{\rule{0.500pt}{0.119pt}}
\multiput(961.00,594.17)(11.962,-13.000){2}{\rule{0.250pt}{0.400pt}}
\multiput(974.58,579.80)(0.493,-0.536){23}{\rule{0.119pt}{0.531pt}}
\multiput(973.17,580.90)(13.000,-12.898){2}{\rule{0.400pt}{0.265pt}}
\multiput(987.58,565.80)(0.493,-0.536){23}{\rule{0.119pt}{0.531pt}}
\multiput(986.17,566.90)(13.000,-12.898){2}{\rule{0.400pt}{0.265pt}}
\multiput(1000.00,552.92)(0.536,-0.493){23}{\rule{0.531pt}{0.119pt}}
\multiput(1000.00,553.17)(12.898,-13.000){2}{\rule{0.265pt}{0.400pt}}
\multiput(1014.58,538.80)(0.493,-0.536){23}{\rule{0.119pt}{0.531pt}}
\multiput(1013.17,539.90)(13.000,-12.898){2}{\rule{0.400pt}{0.265pt}}
\multiput(1027.58,524.80)(0.493,-0.536){23}{\rule{0.119pt}{0.531pt}}
\multiput(1026.17,525.90)(13.000,-12.898){2}{\rule{0.400pt}{0.265pt}}
\multiput(1040.00,511.92)(0.497,-0.493){23}{\rule{0.500pt}{0.119pt}}
\multiput(1040.00,512.17)(11.962,-13.000){2}{\rule{0.250pt}{0.400pt}}
\multiput(1053.00,498.92)(0.536,-0.493){23}{\rule{0.531pt}{0.119pt}}
\multiput(1053.00,499.17)(12.898,-13.000){2}{\rule{0.265pt}{0.400pt}}
\multiput(1067.00,485.92)(0.497,-0.493){23}{\rule{0.500pt}{0.119pt}}
\multiput(1067.00,486.17)(11.962,-13.000){2}{\rule{0.250pt}{0.400pt}}
\multiput(1080.00,472.92)(0.497,-0.493){23}{\rule{0.500pt}{0.119pt}}
\multiput(1080.00,473.17)(11.962,-13.000){2}{\rule{0.250pt}{0.400pt}}
\multiput(1093.00,459.92)(0.539,-0.492){21}{\rule{0.533pt}{0.119pt}}
\multiput(1093.00,460.17)(11.893,-12.000){2}{\rule{0.267pt}{0.400pt}}
\multiput(1106.00,447.92)(0.590,-0.492){19}{\rule{0.573pt}{0.118pt}}
\multiput(1106.00,448.17)(11.811,-11.000){2}{\rule{0.286pt}{0.400pt}}
\multiput(1119.00,436.92)(0.637,-0.492){19}{\rule{0.609pt}{0.118pt}}
\multiput(1119.00,437.17)(12.736,-11.000){2}{\rule{0.305pt}{0.400pt}}
\multiput(1133.00,425.92)(0.590,-0.492){19}{\rule{0.573pt}{0.118pt}}
\multiput(1133.00,426.17)(11.811,-11.000){2}{\rule{0.286pt}{0.400pt}}
\multiput(1146.00,414.92)(0.652,-0.491){17}{\rule{0.620pt}{0.118pt}}
\multiput(1146.00,415.17)(11.713,-10.000){2}{\rule{0.310pt}{0.400pt}}
\multiput(1159.00,404.93)(0.728,-0.489){15}{\rule{0.678pt}{0.118pt}}
\multiput(1159.00,405.17)(11.593,-9.000){2}{\rule{0.339pt}{0.400pt}}
\multiput(1172.00,395.93)(0.786,-0.489){15}{\rule{0.722pt}{0.118pt}}
\multiput(1172.00,396.17)(12.501,-9.000){2}{\rule{0.361pt}{0.400pt}}
\multiput(1186.00,386.93)(0.824,-0.488){13}{\rule{0.750pt}{0.117pt}}
\multiput(1186.00,387.17)(11.443,-8.000){2}{\rule{0.375pt}{0.400pt}}
\multiput(1199.00,378.93)(0.824,-0.488){13}{\rule{0.750pt}{0.117pt}}
\multiput(1199.00,379.17)(11.443,-8.000){2}{\rule{0.375pt}{0.400pt}}
\multiput(1212.00,370.93)(0.950,-0.485){11}{\rule{0.843pt}{0.117pt}}
\multiput(1212.00,371.17)(11.251,-7.000){2}{\rule{0.421pt}{0.400pt}}
\multiput(1225.00,363.93)(1.214,-0.482){9}{\rule{1.033pt}{0.116pt}}
\multiput(1225.00,364.17)(11.855,-6.000){2}{\rule{0.517pt}{0.400pt}}
\multiput(1239.00,357.93)(1.123,-0.482){9}{\rule{0.967pt}{0.116pt}}
\multiput(1239.00,358.17)(10.994,-6.000){2}{\rule{0.483pt}{0.400pt}}
\multiput(1252.00,351.93)(1.378,-0.477){7}{\rule{1.140pt}{0.115pt}}
\multiput(1252.00,352.17)(10.634,-5.000){2}{\rule{0.570pt}{0.400pt}}
\multiput(1265.00,346.93)(1.378,-0.477){7}{\rule{1.140pt}{0.115pt}}
\multiput(1265.00,347.17)(10.634,-5.000){2}{\rule{0.570pt}{0.400pt}}
\multiput(1278.00,341.93)(1.489,-0.477){7}{\rule{1.220pt}{0.115pt}}
\multiput(1278.00,342.17)(11.468,-5.000){2}{\rule{0.610pt}{0.400pt}}
\multiput(1292.00,336.94)(1.797,-0.468){5}{\rule{1.400pt}{0.113pt}}
\multiput(1292.00,337.17)(10.094,-4.000){2}{\rule{0.700pt}{0.400pt}}
\multiput(1305.00,332.95)(2.695,-0.447){3}{\rule{1.833pt}{0.108pt}}
\multiput(1305.00,333.17)(9.195,-3.000){2}{\rule{0.917pt}{0.400pt}}
\multiput(1318.00,329.95)(2.695,-0.447){3}{\rule{1.833pt}{0.108pt}}
\multiput(1318.00,330.17)(9.195,-3.000){2}{\rule{0.917pt}{0.400pt}}
\multiput(1331.00,326.95)(2.918,-0.447){3}{\rule{1.967pt}{0.108pt}}
\multiput(1331.00,327.17)(9.918,-3.000){2}{\rule{0.983pt}{0.400pt}}
\put(1345,323.17){\rule{2.700pt}{0.400pt}}
\multiput(1345.00,324.17)(7.396,-2.000){2}{\rule{1.350pt}{0.400pt}}
\put(1358,321.17){\rule{2.700pt}{0.400pt}}
\multiput(1358.00,322.17)(7.396,-2.000){2}{\rule{1.350pt}{0.400pt}}
\put(1371,319.67){\rule{3.132pt}{0.400pt}}
\multiput(1371.00,320.17)(6.500,-1.000){2}{\rule{1.566pt}{0.400pt}}
\put(1384,318.67){\rule{3.373pt}{0.400pt}}
\multiput(1384.00,319.17)(7.000,-1.000){2}{\rule{1.686pt}{0.400pt}}
\put(1398,317.67){\rule{3.132pt}{0.400pt}}
\multiput(1398.00,318.17)(6.500,-1.000){2}{\rule{1.566pt}{0.400pt}}
\put(100.0,318.0){\rule[-0.200pt]{3.132pt}{0.400pt}}
\put(1411.0,318.0){\rule[-0.200pt]{3.132pt}{0.400pt}}
\put(100,123){\usebox{\plotpoint}}
\put(113,122.67){\rule{3.132pt}{0.400pt}}
\multiput(113.00,122.17)(6.500,1.000){2}{\rule{1.566pt}{0.400pt}}
\put(126,123.67){\rule{3.373pt}{0.400pt}}
\multiput(126.00,123.17)(7.000,1.000){2}{\rule{1.686pt}{0.400pt}}
\put(140,124.67){\rule{3.132pt}{0.400pt}}
\multiput(140.00,124.17)(6.500,1.000){2}{\rule{1.566pt}{0.400pt}}
\put(153,126.17){\rule{2.700pt}{0.400pt}}
\multiput(153.00,125.17)(7.396,2.000){2}{\rule{1.350pt}{0.400pt}}
\put(166,128.17){\rule{2.700pt}{0.400pt}}
\multiput(166.00,127.17)(7.396,2.000){2}{\rule{1.350pt}{0.400pt}}
\multiput(179.00,130.61)(2.918,0.447){3}{\rule{1.967pt}{0.108pt}}
\multiput(179.00,129.17)(9.918,3.000){2}{\rule{0.983pt}{0.400pt}}
\multiput(193.00,133.61)(2.695,0.447){3}{\rule{1.833pt}{0.108pt}}
\multiput(193.00,132.17)(9.195,3.000){2}{\rule{0.917pt}{0.400pt}}
\multiput(206.00,136.60)(1.797,0.468){5}{\rule{1.400pt}{0.113pt}}
\multiput(206.00,135.17)(10.094,4.000){2}{\rule{0.700pt}{0.400pt}}
\multiput(219.00,140.60)(1.797,0.468){5}{\rule{1.400pt}{0.113pt}}
\multiput(219.00,139.17)(10.094,4.000){2}{\rule{0.700pt}{0.400pt}}
\multiput(232.00,144.59)(1.489,0.477){7}{\rule{1.220pt}{0.115pt}}
\multiput(232.00,143.17)(11.468,5.000){2}{\rule{0.610pt}{0.400pt}}
\multiput(246.00,149.59)(1.123,0.482){9}{\rule{0.967pt}{0.116pt}}
\multiput(246.00,148.17)(10.994,6.000){2}{\rule{0.483pt}{0.400pt}}
\multiput(259.00,155.59)(1.123,0.482){9}{\rule{0.967pt}{0.116pt}}
\multiput(259.00,154.17)(10.994,6.000){2}{\rule{0.483pt}{0.400pt}}
\multiput(272.00,161.59)(0.950,0.485){11}{\rule{0.843pt}{0.117pt}}
\multiput(272.00,160.17)(11.251,7.000){2}{\rule{0.421pt}{0.400pt}}
\multiput(285.00,168.59)(0.890,0.488){13}{\rule{0.800pt}{0.117pt}}
\multiput(285.00,167.17)(12.340,8.000){2}{\rule{0.400pt}{0.400pt}}
\multiput(299.00,176.59)(0.728,0.489){15}{\rule{0.678pt}{0.118pt}}
\multiput(299.00,175.17)(11.593,9.000){2}{\rule{0.339pt}{0.400pt}}
\multiput(312.00,185.59)(0.728,0.489){15}{\rule{0.678pt}{0.118pt}}
\multiput(312.00,184.17)(11.593,9.000){2}{\rule{0.339pt}{0.400pt}}
\multiput(325.00,194.58)(0.590,0.492){19}{\rule{0.573pt}{0.118pt}}
\multiput(325.00,193.17)(11.811,11.000){2}{\rule{0.286pt}{0.400pt}}
\multiput(338.00,205.58)(0.582,0.492){21}{\rule{0.567pt}{0.119pt}}
\multiput(338.00,204.17)(12.824,12.000){2}{\rule{0.283pt}{0.400pt}}
\multiput(352.58,217.00)(0.493,0.536){23}{\rule{0.119pt}{0.531pt}}
\multiput(351.17,217.00)(13.000,12.898){2}{\rule{0.400pt}{0.265pt}}
\multiput(365.58,231.00)(0.493,0.536){23}{\rule{0.119pt}{0.531pt}}
\multiput(364.17,231.00)(13.000,12.898){2}{\rule{0.400pt}{0.265pt}}
\multiput(378.58,245.00)(0.493,0.616){23}{\rule{0.119pt}{0.592pt}}
\multiput(377.17,245.00)(13.000,14.771){2}{\rule{0.400pt}{0.296pt}}
\multiput(391.58,261.00)(0.494,0.644){25}{\rule{0.119pt}{0.614pt}}
\multiput(390.17,261.00)(14.000,16.725){2}{\rule{0.400pt}{0.307pt}}
\multiput(405.58,279.00)(0.493,0.734){23}{\rule{0.119pt}{0.685pt}}
\multiput(404.17,279.00)(13.000,17.579){2}{\rule{0.400pt}{0.342pt}}
\multiput(418.58,298.00)(0.493,0.774){23}{\rule{0.119pt}{0.715pt}}
\multiput(417.17,298.00)(13.000,18.515){2}{\rule{0.400pt}{0.358pt}}
\multiput(431.58,318.00)(0.493,0.853){23}{\rule{0.119pt}{0.777pt}}
\multiput(430.17,318.00)(13.000,20.387){2}{\rule{0.400pt}{0.388pt}}
\multiput(444.58,340.00)(0.493,0.933){23}{\rule{0.119pt}{0.838pt}}
\multiput(443.17,340.00)(13.000,22.260){2}{\rule{0.400pt}{0.419pt}}
\multiput(457.58,364.00)(0.494,0.901){25}{\rule{0.119pt}{0.814pt}}
\multiput(456.17,364.00)(14.000,23.310){2}{\rule{0.400pt}{0.407pt}}
\multiput(471.58,389.00)(0.493,1.012){23}{\rule{0.119pt}{0.900pt}}
\multiput(470.17,389.00)(13.000,24.132){2}{\rule{0.400pt}{0.450pt}}
\multiput(484.58,415.00)(0.493,1.091){23}{\rule{0.119pt}{0.962pt}}
\multiput(483.17,415.00)(13.000,26.004){2}{\rule{0.400pt}{0.481pt}}
\multiput(497.58,443.00)(0.493,1.131){23}{\rule{0.119pt}{0.992pt}}
\multiput(496.17,443.00)(13.000,26.940){2}{\rule{0.400pt}{0.496pt}}
\multiput(510.58,472.00)(0.494,1.084){25}{\rule{0.119pt}{0.957pt}}
\multiput(509.17,472.00)(14.000,28.013){2}{\rule{0.400pt}{0.479pt}}
\multiput(524.58,502.00)(0.493,1.210){23}{\rule{0.119pt}{1.054pt}}
\multiput(523.17,502.00)(13.000,28.813){2}{\rule{0.400pt}{0.527pt}}
\multiput(537.58,533.00)(0.493,1.210){23}{\rule{0.119pt}{1.054pt}}
\multiput(536.17,533.00)(13.000,28.813){2}{\rule{0.400pt}{0.527pt}}
\multiput(550.58,564.00)(0.493,1.250){23}{\rule{0.119pt}{1.085pt}}
\multiput(549.17,564.00)(13.000,29.749){2}{\rule{0.400pt}{0.542pt}}
\multiput(563.58,596.00)(0.494,1.121){25}{\rule{0.119pt}{0.986pt}}
\multiput(562.17,596.00)(14.000,28.954){2}{\rule{0.400pt}{0.493pt}}
\multiput(577.58,627.00)(0.493,1.250){23}{\rule{0.119pt}{1.085pt}}
\multiput(576.17,627.00)(13.000,29.749){2}{\rule{0.400pt}{0.542pt}}
\multiput(590.58,659.00)(0.493,1.171){23}{\rule{0.119pt}{1.023pt}}
\multiput(589.17,659.00)(13.000,27.877){2}{\rule{0.400pt}{0.512pt}}
\multiput(603.58,689.00)(0.493,1.171){23}{\rule{0.119pt}{1.023pt}}
\multiput(602.17,689.00)(13.000,27.877){2}{\rule{0.400pt}{0.512pt}}
\multiput(616.58,719.00)(0.494,1.048){25}{\rule{0.119pt}{0.929pt}}
\multiput(615.17,719.00)(14.000,27.073){2}{\rule{0.400pt}{0.464pt}}
\multiput(630.58,748.00)(0.493,1.052){23}{\rule{0.119pt}{0.931pt}}
\multiput(629.17,748.00)(13.000,25.068){2}{\rule{0.400pt}{0.465pt}}
\multiput(643.58,775.00)(0.493,0.972){23}{\rule{0.119pt}{0.869pt}}
\multiput(642.17,775.00)(13.000,23.196){2}{\rule{0.400pt}{0.435pt}}
\multiput(656.58,800.00)(0.493,0.893){23}{\rule{0.119pt}{0.808pt}}
\multiput(655.17,800.00)(13.000,21.324){2}{\rule{0.400pt}{0.404pt}}
\multiput(669.58,823.00)(0.494,0.754){25}{\rule{0.119pt}{0.700pt}}
\multiput(668.17,823.00)(14.000,19.547){2}{\rule{0.400pt}{0.350pt}}
\multiput(683.58,844.00)(0.493,0.695){23}{\rule{0.119pt}{0.654pt}}
\multiput(682.17,844.00)(13.000,16.643){2}{\rule{0.400pt}{0.327pt}}
\multiput(696.58,862.00)(0.493,0.576){23}{\rule{0.119pt}{0.562pt}}
\multiput(695.17,862.00)(13.000,13.834){2}{\rule{0.400pt}{0.281pt}}
\multiput(709.00,877.58)(0.539,0.492){21}{\rule{0.533pt}{0.119pt}}
\multiput(709.00,876.17)(11.893,12.000){2}{\rule{0.267pt}{0.400pt}}
\multiput(722.00,889.59)(0.890,0.488){13}{\rule{0.800pt}{0.117pt}}
\multiput(722.00,888.17)(12.340,8.000){2}{\rule{0.400pt}{0.400pt}}
\multiput(736.00,897.59)(1.378,0.477){7}{\rule{1.140pt}{0.115pt}}
\multiput(736.00,896.17)(10.634,5.000){2}{\rule{0.570pt}{0.400pt}}
\put(749,902.17){\rule{2.700pt}{0.400pt}}
\multiput(749.00,901.17)(7.396,2.000){2}{\rule{1.350pt}{0.400pt}}
\put(762,902.17){\rule{2.700pt}{0.400pt}}
\multiput(762.00,903.17)(7.396,-2.000){2}{\rule{1.350pt}{0.400pt}}
\multiput(775.00,900.93)(1.378,-0.477){7}{\rule{1.140pt}{0.115pt}}
\multiput(775.00,901.17)(10.634,-5.000){2}{\rule{0.570pt}{0.400pt}}
\multiput(788.00,895.93)(0.890,-0.488){13}{\rule{0.800pt}{0.117pt}}
\multiput(788.00,896.17)(12.340,-8.000){2}{\rule{0.400pt}{0.400pt}}
\multiput(802.00,887.92)(0.539,-0.492){21}{\rule{0.533pt}{0.119pt}}
\multiput(802.00,888.17)(11.893,-12.000){2}{\rule{0.267pt}{0.400pt}}
\multiput(815.58,874.67)(0.493,-0.576){23}{\rule{0.119pt}{0.562pt}}
\multiput(814.17,875.83)(13.000,-13.834){2}{\rule{0.400pt}{0.281pt}}
\multiput(828.58,859.29)(0.493,-0.695){23}{\rule{0.119pt}{0.654pt}}
\multiput(827.17,860.64)(13.000,-16.643){2}{\rule{0.400pt}{0.327pt}}
\multiput(841.58,841.09)(0.494,-0.754){25}{\rule{0.119pt}{0.700pt}}
\multiput(840.17,842.55)(14.000,-19.547){2}{\rule{0.400pt}{0.350pt}}
\multiput(855.58,819.65)(0.493,-0.893){23}{\rule{0.119pt}{0.808pt}}
\multiput(854.17,821.32)(13.000,-21.324){2}{\rule{0.400pt}{0.404pt}}
\multiput(868.58,796.39)(0.493,-0.972){23}{\rule{0.119pt}{0.869pt}}
\multiput(867.17,798.20)(13.000,-23.196){2}{\rule{0.400pt}{0.435pt}}
\multiput(881.58,771.14)(0.493,-1.052){23}{\rule{0.119pt}{0.931pt}}
\multiput(880.17,773.07)(13.000,-25.068){2}{\rule{0.400pt}{0.465pt}}
\multiput(894.58,744.15)(0.494,-1.048){25}{\rule{0.119pt}{0.929pt}}
\multiput(893.17,746.07)(14.000,-27.073){2}{\rule{0.400pt}{0.464pt}}
\multiput(908.58,714.75)(0.493,-1.171){23}{\rule{0.119pt}{1.023pt}}
\multiput(907.17,716.88)(13.000,-27.877){2}{\rule{0.400pt}{0.512pt}}
\multiput(921.58,684.75)(0.493,-1.171){23}{\rule{0.119pt}{1.023pt}}
\multiput(920.17,686.88)(13.000,-27.877){2}{\rule{0.400pt}{0.512pt}}
\multiput(934.58,654.50)(0.493,-1.250){23}{\rule{0.119pt}{1.085pt}}
\multiput(933.17,656.75)(13.000,-29.749){2}{\rule{0.400pt}{0.542pt}}
\multiput(947.58,622.91)(0.494,-1.121){25}{\rule{0.119pt}{0.986pt}}
\multiput(946.17,624.95)(14.000,-28.954){2}{\rule{0.400pt}{0.493pt}}
\multiput(961.58,591.50)(0.493,-1.250){23}{\rule{0.119pt}{1.085pt}}
\multiput(960.17,593.75)(13.000,-29.749){2}{\rule{0.400pt}{0.542pt}}
\multiput(974.58,559.63)(0.493,-1.210){23}{\rule{0.119pt}{1.054pt}}
\multiput(973.17,561.81)(13.000,-28.813){2}{\rule{0.400pt}{0.527pt}}
\multiput(987.58,528.63)(0.493,-1.210){23}{\rule{0.119pt}{1.054pt}}
\multiput(986.17,530.81)(13.000,-28.813){2}{\rule{0.400pt}{0.527pt}}
\multiput(1000.58,498.03)(0.494,-1.084){25}{\rule{0.119pt}{0.957pt}}
\multiput(999.17,500.01)(14.000,-28.013){2}{\rule{0.400pt}{0.479pt}}
\multiput(1014.58,467.88)(0.493,-1.131){23}{\rule{0.119pt}{0.992pt}}
\multiput(1013.17,469.94)(13.000,-26.940){2}{\rule{0.400pt}{0.496pt}}
\multiput(1027.58,439.01)(0.493,-1.091){23}{\rule{0.119pt}{0.962pt}}
\multiput(1026.17,441.00)(13.000,-26.004){2}{\rule{0.400pt}{0.481pt}}
\multiput(1040.58,411.26)(0.493,-1.012){23}{\rule{0.119pt}{0.900pt}}
\multiput(1039.17,413.13)(13.000,-24.132){2}{\rule{0.400pt}{0.450pt}}
\multiput(1053.58,385.62)(0.494,-0.901){25}{\rule{0.119pt}{0.814pt}}
\multiput(1052.17,387.31)(14.000,-23.310){2}{\rule{0.400pt}{0.407pt}}
\multiput(1067.58,360.52)(0.493,-0.933){23}{\rule{0.119pt}{0.838pt}}
\multiput(1066.17,362.26)(13.000,-22.260){2}{\rule{0.400pt}{0.419pt}}
\multiput(1080.58,336.77)(0.493,-0.853){23}{\rule{0.119pt}{0.777pt}}
\multiput(1079.17,338.39)(13.000,-20.387){2}{\rule{0.400pt}{0.388pt}}
\multiput(1093.58,315.03)(0.493,-0.774){23}{\rule{0.119pt}{0.715pt}}
\multiput(1092.17,316.52)(13.000,-18.515){2}{\rule{0.400pt}{0.358pt}}
\multiput(1106.58,295.16)(0.493,-0.734){23}{\rule{0.119pt}{0.685pt}}
\multiput(1105.17,296.58)(13.000,-17.579){2}{\rule{0.400pt}{0.342pt}}
\multiput(1119.58,276.45)(0.494,-0.644){25}{\rule{0.119pt}{0.614pt}}
\multiput(1118.17,277.73)(14.000,-16.725){2}{\rule{0.400pt}{0.307pt}}
\multiput(1133.58,258.54)(0.493,-0.616){23}{\rule{0.119pt}{0.592pt}}
\multiput(1132.17,259.77)(13.000,-14.771){2}{\rule{0.400pt}{0.296pt}}
\multiput(1146.58,242.80)(0.493,-0.536){23}{\rule{0.119pt}{0.531pt}}
\multiput(1145.17,243.90)(13.000,-12.898){2}{\rule{0.400pt}{0.265pt}}
\multiput(1159.58,228.80)(0.493,-0.536){23}{\rule{0.119pt}{0.531pt}}
\multiput(1158.17,229.90)(13.000,-12.898){2}{\rule{0.400pt}{0.265pt}}
\multiput(1172.00,215.92)(0.582,-0.492){21}{\rule{0.567pt}{0.119pt}}
\multiput(1172.00,216.17)(12.824,-12.000){2}{\rule{0.283pt}{0.400pt}}
\multiput(1186.00,203.92)(0.590,-0.492){19}{\rule{0.573pt}{0.118pt}}
\multiput(1186.00,204.17)(11.811,-11.000){2}{\rule{0.286pt}{0.400pt}}
\multiput(1199.00,192.93)(0.728,-0.489){15}{\rule{0.678pt}{0.118pt}}
\multiput(1199.00,193.17)(11.593,-9.000){2}{\rule{0.339pt}{0.400pt}}
\multiput(1212.00,183.93)(0.728,-0.489){15}{\rule{0.678pt}{0.118pt}}
\multiput(1212.00,184.17)(11.593,-9.000){2}{\rule{0.339pt}{0.400pt}}
\multiput(1225.00,174.93)(0.890,-0.488){13}{\rule{0.800pt}{0.117pt}}
\multiput(1225.00,175.17)(12.340,-8.000){2}{\rule{0.400pt}{0.400pt}}
\multiput(1239.00,166.93)(0.950,-0.485){11}{\rule{0.843pt}{0.117pt}}
\multiput(1239.00,167.17)(11.251,-7.000){2}{\rule{0.421pt}{0.400pt}}
\multiput(1252.00,159.93)(1.123,-0.482){9}{\rule{0.967pt}{0.116pt}}
\multiput(1252.00,160.17)(10.994,-6.000){2}{\rule{0.483pt}{0.400pt}}
\multiput(1265.00,153.93)(1.123,-0.482){9}{\rule{0.967pt}{0.116pt}}
\multiput(1265.00,154.17)(10.994,-6.000){2}{\rule{0.483pt}{0.400pt}}
\multiput(1278.00,147.93)(1.489,-0.477){7}{\rule{1.220pt}{0.115pt}}
\multiput(1278.00,148.17)(11.468,-5.000){2}{\rule{0.610pt}{0.400pt}}
\multiput(1292.00,142.94)(1.797,-0.468){5}{\rule{1.400pt}{0.113pt}}
\multiput(1292.00,143.17)(10.094,-4.000){2}{\rule{0.700pt}{0.400pt}}
\multiput(1305.00,138.94)(1.797,-0.468){5}{\rule{1.400pt}{0.113pt}}
\multiput(1305.00,139.17)(10.094,-4.000){2}{\rule{0.700pt}{0.400pt}}
\multiput(1318.00,134.95)(2.695,-0.447){3}{\rule{1.833pt}{0.108pt}}
\multiput(1318.00,135.17)(9.195,-3.000){2}{\rule{0.917pt}{0.400pt}}
\multiput(1331.00,131.95)(2.918,-0.447){3}{\rule{1.967pt}{0.108pt}}
\multiput(1331.00,132.17)(9.918,-3.000){2}{\rule{0.983pt}{0.400pt}}
\put(1345,128.17){\rule{2.700pt}{0.400pt}}
\multiput(1345.00,129.17)(7.396,-2.000){2}{\rule{1.350pt}{0.400pt}}
\put(1358,126.17){\rule{2.700pt}{0.400pt}}
\multiput(1358.00,127.17)(7.396,-2.000){2}{\rule{1.350pt}{0.400pt}}
\put(1371,124.67){\rule{3.132pt}{0.400pt}}
\multiput(1371.00,125.17)(6.500,-1.000){2}{\rule{1.566pt}{0.400pt}}
\put(1384,123.67){\rule{3.373pt}{0.400pt}}
\multiput(1384.00,124.17)(7.000,-1.000){2}{\rule{1.686pt}{0.400pt}}
\put(1398,122.67){\rule{3.132pt}{0.400pt}}
\multiput(1398.00,123.17)(6.500,-1.000){2}{\rule{1.566pt}{0.400pt}}
\put(100.0,123.0){\rule[-0.200pt]{3.132pt}{0.400pt}}
\put(1411.0,123.0){\rule[-0.200pt]{3.132pt}{0.400pt}}
\end{picture}
\vspace{0.5cm}
\caption{Behaviors of $z$ and $R$ changing with $\theta$ under the
condition $|A^{+-}| =|A^{+0}|$.}
\end{figure}

\section{Direct $CP$ asymmetries}
 
We proceed with the factorization scheme to calculate direct
$CP$ asymmetries in the decay modes under discussion. 
As for the final states with two vector mesons, we sum over 
their polarizations and arrive at 
$|\bar{Z}|^2 = |Z|^2$, a relationship which apparently holds 
for other types of final states.
With the help of Eqs. (6)
and (7) it is easy to show that the decay rate asymmetry between
$B^+_u\rightarrow D^{(*)+}\bar{D}^{(*)0}$ and
$B^-_u\rightarrow D^{(*)-}D^{(*)0}$ decays is identical to that
between $B^0_d$ and $\bar{B}^0_d \rightarrow D^{(*)+}D^{(*)-}$
or $D^{(*)0}\bar{D}^{(*)0}$ decays. All these $CP$ asymmetries are
independent of the rescattering phases 
and the hadronic matrix elements
\footnote{For time-integrated $B_d$ decays the 
direct $CP$ asymmetries are diluted by a well-known
factor $1/(1+x^2_d)$, where $x_d \approx 0.7$ is the
$B^0_d$-$\bar{B}^0_d$ mixing parameter. In this paper we do not
take such mixing effects into account.}:
\begin{eqnarray}
{\cal A} & = & \frac{|\bar{A}^{-0}|^2 -
|A^{+0}|^2}{|\bar{A}^{-0}|^2 + |A^{+0}|^2} 
\nonumber \\
& = & \frac{|\bar{A}^{+-}|^2 - |A^{+-}|^2}
{|\bar{A}^{+-}|^2 + |A^{+-}|^2} 
\nonumber \\
& = & \frac{|\bar{A}^{00}|^2 - |A^{00}|^2}
{|\bar{A}^{00}|^2 + |A^{00}|^2} 
\nonumber \\
& = & \frac{2r \sin\gamma {\rm Im}(\zeta_u \zeta^*_c)}
{r^2 |\zeta_u|^2 + |\zeta_c|^2 -2r \cos\gamma {\rm Re}
(\zeta_u \zeta^*_c)} \; ,
\end{eqnarray}
where $r$ and $\gamma$ are defined by
$r e^{{\rm i}\gamma} \equiv - (V_{ud}V^*_{ub})/(V_{cd}V^*_{cb})$.
The phase $\gamma$ corresponds to another inner angle 
of the unitarity triangle defined in Eq. (1).
Eq. (10) indicates that direct $CP$ violation arises 
only from final-state interactions of the quark level
(through the penguin mechanism) in $B\rightarrow
D^{(*)}\bar{D}^{(*)}$ decays. This result, as a straightforward
consequence of the factorization approximation, can
directly be confronted with the upcoming experiments at $B$-meson
factories.

\vspace{0.4cm}

Let us evaluate the direct $CP$ asymmetries ${\cal A}$
for different final states. As mentioned above, $S_u$ and
$S_c$ in Eq. (10) depend on the effective Wilson coefficients
$\bar{c}_i$ and the penguin loop-integral functions $F_q$. 
The latter can be given, for a momentum-squared transfer $k^2$ at the
$O(m_b)$ scale, as follows \cite{Soni}:
\begin{equation}
F_q \; =\; 4 \int^1_0 {\rm d}x ~ x (1-x) \ln \left [
\frac{m^2_q - k^2 x (1-x)}{m^2_b} \right ] \; .
\end{equation}
The absorptive part of $F_q$, which is a necessary condition for
direct $CP$ violation, emerges if $k^2 \geq 4m^2_q$.
The concrete 
expressions of $S_u$ and $S_c$ are found to be
\begin{eqnarray}
S_u & = & C_{\bf 1} + C_{\bf 3} + C_{\bf 4}
\frac{1+\xi}{9\pi} \left (\frac{10}{3} + F_u \right ) \; ,
\nonumber \\
S_c & = & C_{\bf 2} + C_{\bf 3} + C_{\bf 4}
\frac{1+\xi}{9\pi} \left (\frac{10}{3} + F_c \right ) \; ,
\end{eqnarray}
where 
\begin{eqnarray}
C_{\bf 1} & = & \frac{\bar{c}_3}{3} + \bar{c}_4 + \frac{\bar{c}_9}{3}
+ \bar{c}_{10} \; , \nonumber \\
C_{\bf 2} & = & \frac{\bar{c}_1}{3} + \bar{c}_2 + C_{\bf 1} \; ,
\nonumber \\
C_{\bf 3} & = & \frac{\bar{c}_5}{3} + \bar{c}_6 + \frac{\bar{c}_7}{3}
+ \bar{c}_8 \; , \nonumber \\
C_{\bf 4} & = & \bar{c}_2 \alpha_s + \left (\bar{c}_1 +
\frac{\bar{c}_2}{3} \right ) \alpha_e \; .
\end{eqnarray}
In these equations $\bar{c}_i$ (for $i=1, \cdot\cdot\cdot, 10$) are the
renormalization-scheme-independent Wilson coefficients, 
$\alpha_s$ and $\alpha_e$ stand respectively for the strong and
electroweak coupling constants, and $\xi$ is a factorization
parameter arising from the transformation of (V--A)(V+A) currents
into (V--A)(V--A) ones for the penguin operators $Q_5, \cdot\cdot\cdot, Q_8$.
Note that $\xi$ depends on properties of the final-state mesons
\cite{Xing93}:
\begin{equation}
\xi \; =\; \left \{ \matrix{
+ \displaystyle\frac{2 m^2_D }{(m_c + m_d) (m_b - m_c)} & (D\bar{D}) \cr
0 & (D^* \bar{D}) \cr
- \displaystyle\frac{2 m^2_D }{(m_c + m_d) (m_b + m_c)} & (D\bar{D}^*) \cr
0 & (D^* \bar{D}^*) \cr} \; , \right .
\end{equation}
where the order of two $D^{(*)}$ mesons
corresponds to that in the factorized hadronic 
matrix element $Z$ or $\bar{Z}$, as given in Eq. (5).

\vspace{0.4cm}

With the help of Eqs. (11) -- (14) we are able to calculate the $CP$
asymmetries ${\cal A}$ numerically.
Note that $|S_u| \ll |S_c|$, as
the former consists only of the penguin contribution and the latter is
dominated by the much larger tree-level contribution. This, together
with $|r| <1$, allows an instructive analytical approximation of 
${\cal A}$:
\begin{equation}
{\cal A} \; \approx \; 2r \sin\gamma {\rm Im}
\left ( \frac{S_u}{S_c} \right ) \; .
\end{equation}
For illustration, we typically choose
$m_u =5$ MeV, $m_c =1.35$ GeV, $m_b =5$ GeV and $m_t =174$ GeV.
The strong coupling constant is taken as 
$\alpha_s =0.21$ at the $O(m_b)$ scale.
Values of the effective coefficients $\bar{c}_i$ read \cite{He}:
$\bar{c}_1 = -0.313$, $\bar{c}_2 =1.150$, $\bar{c}_3 =0.017$,
$\bar{c}_4 =-0.037$, $\bar{c}_5 =0.010$, $\bar{c}_6 =-0.046$,
$\bar{c}_7 =-0.001\alpha_e$, $\bar{c}_8 =0.049\alpha_e$,
$\bar{c}_9 =-1.321\alpha_e$ and $\bar{c}_{10} =0.267 \alpha_e$
with $\alpha_e =1/128$. The CKM factors are taken to be
$r=0.38$ and $\gamma =60^{\circ}$, consistent with the lastest
data on quark mixing and $CP$ violation \cite{Stocchi}. 
The unknown penguin
momentum transfer $k^2$ is treated as a free parameter changing
from $0.01m^2_b$ to $m^2_b$. Our numerical results are shown
in Fig. 2. Some discussions are in order.
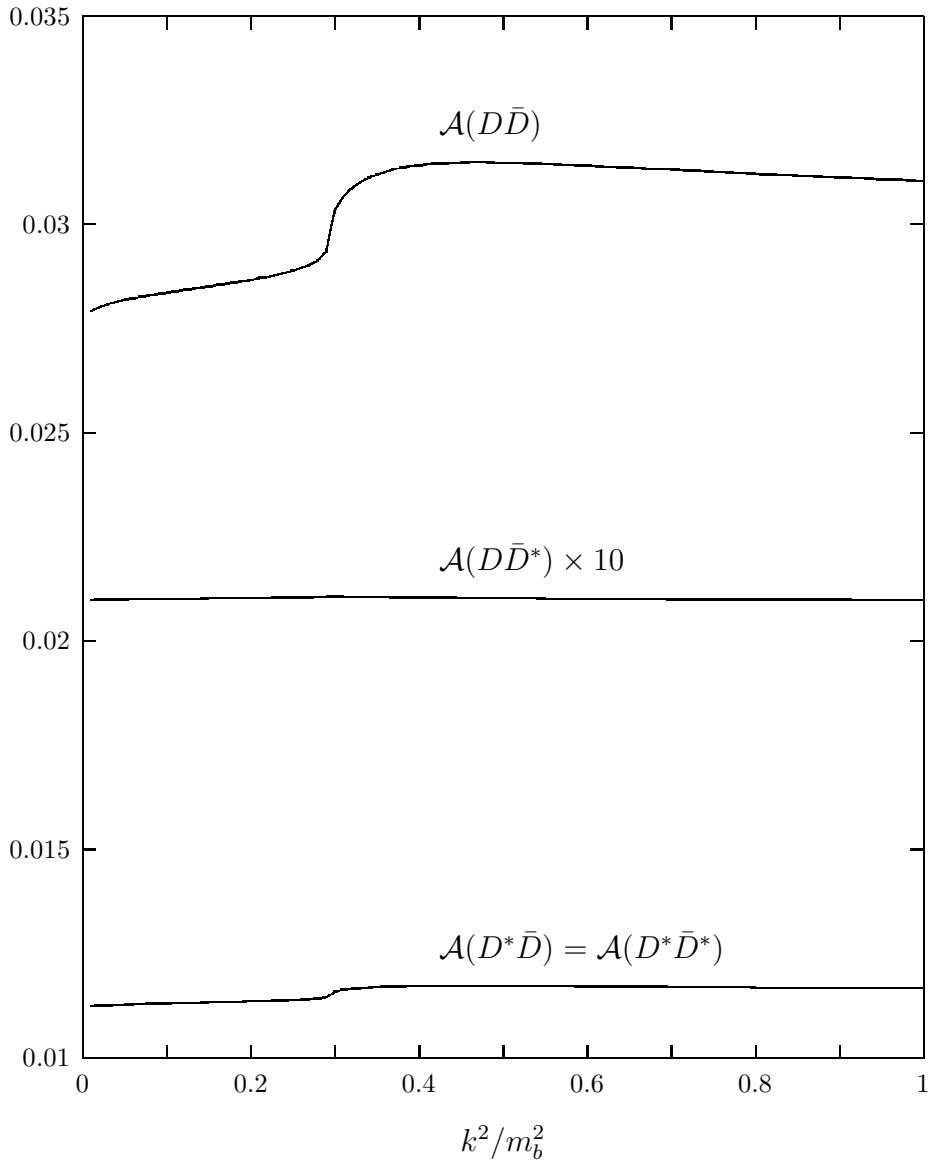
\begin{figure}
\setlength{\unitlength}{0.240900pt}
\ifx\plotpoint\undefined\newsavebox{\plotpoint}\fi
\sbox{\plotpoint}{\rule[-0.200pt]{0.400pt}{0.400pt}}%
\begin{picture}(1500,1800)(-250,0)
\font\gnuplot=cmr10 at 10pt
\gnuplot
\sbox{\plotpoint}{\rule[-0.200pt]{0.400pt}{0.400pt}}%
\put(140.0,123.0){\rule[-0.200pt]{4.818pt}{0.400pt}}
\put(120,123){\makebox(0,0)[r]{0.01}}
\put(1440.0,123.0){\rule[-0.200pt]{4.818pt}{0.400pt}}
\put(140.0,450.0){\rule[-0.200pt]{4.818pt}{0.400pt}}
\put(120,450){\makebox(0,0)[r]{0.015}}
\put(1440.0,450.0){\rule[-0.200pt]{4.818pt}{0.400pt}}
\put(140.0,778.0){\rule[-0.200pt]{4.818pt}{0.400pt}}
\put(120,778){\makebox(0,0)[r]{0.02}}
\put(1440.0,778.0){\rule[-0.200pt]{4.818pt}{0.400pt}}
\put(140.0,1105.0){\rule[-0.200pt]{4.818pt}{0.400pt}}
\put(120,1105){\makebox(0,0)[r]{0.025}}
\put(1440.0,1105.0){\rule[-0.200pt]{4.818pt}{0.400pt}}
\put(140.0,1433.0){\rule[-0.200pt]{4.818pt}{0.400pt}}
\put(120,1433){\makebox(0,0)[r]{0.03}}
\put(1440.0,1433.0){\rule[-0.200pt]{4.818pt}{0.400pt}}
\put(140.0,1760.0){\rule[-0.200pt]{4.818pt}{0.400pt}}
\put(120,1760){\makebox(0,0)[r]{0.035}}
\put(1440.0,1760.0){\rule[-0.200pt]{4.818pt}{0.400pt}}
\put(140.0,123.0){\rule[-0.200pt]{0.400pt}{4.818pt}}
\put(140,82){\makebox(0,0){0}}
\put(140.0,1740.0){\rule[-0.200pt]{0.400pt}{4.818pt}}
\put(272.0,123.0){\rule[-0.200pt]{0.400pt}{4.818pt}}
\put(272.0,1740.0){\rule[-0.200pt]{0.400pt}{4.818pt}}
\put(404.0,123.0){\rule[-0.200pt]{0.400pt}{4.818pt}}
\put(404,82){\makebox(0,0){0.2}}
\put(404.0,1740.0){\rule[-0.200pt]{0.400pt}{4.818pt}}
\put(536.0,123.0){\rule[-0.200pt]{0.400pt}{4.818pt}}
\put(536.0,1740.0){\rule[-0.200pt]{0.400pt}{4.818pt}}
\put(668.0,123.0){\rule[-0.200pt]{0.400pt}{4.818pt}}
\put(668,82){\makebox(0,0){0.4}}
\put(668.0,1740.0){\rule[-0.200pt]{0.400pt}{4.818pt}}
\put(800.0,123.0){\rule[-0.200pt]{0.400pt}{4.818pt}}
\put(800.0,1740.0){\rule[-0.200pt]{0.400pt}{4.818pt}}
\put(932.0,123.0){\rule[-0.200pt]{0.400pt}{4.818pt}}
\put(932,82){\makebox(0,0){0.6}}
\put(932.0,1740.0){\rule[-0.200pt]{0.400pt}{4.818pt}}
\put(1064.0,123.0){\rule[-0.200pt]{0.400pt}{4.818pt}}
\put(1064.0,1740.0){\rule[-0.200pt]{0.400pt}{4.818pt}}
\put(1196.0,123.0){\rule[-0.200pt]{0.400pt}{4.818pt}}
\put(1196,82){\makebox(0,0){0.8}}
\put(1196.0,1740.0){\rule[-0.200pt]{0.400pt}{4.818pt}}
\put(1328.0,123.0){\rule[-0.200pt]{0.400pt}{4.818pt}}
\put(1328.0,1740.0){\rule[-0.200pt]{0.400pt}{4.818pt}}
\put(1460.0,123.0){\rule[-0.200pt]{0.400pt}{4.818pt}}
\put(1460,82){\makebox(0,0){1}}
\put(1460.0,1740.0){\rule[-0.200pt]{0.400pt}{4.818pt}}
\put(140.0,123.0){\rule[-0.200pt]{317.988pt}{0.400pt}}
\put(1460.0,123.0){\rule[-0.200pt]{0.400pt}{394.353pt}}
\put(140.0,1760.0){\rule[-0.200pt]{317.988pt}{0.400pt}}
\put(800,-10){\makebox(0,0){$k^2/m^2_b$}}
\put(700,1575){$\scriptsize {\cal A}(D\bar{D})$}
\put(700,890){$\scriptsize {\cal A}(D\bar{D}^*) \times 10$}
\put(700,280){$\scriptsize {\cal A}(D^*\bar{D}) =
{\cal A}(D^*\bar{D}^*)$}
\put(140.0,123.0){\rule[-0.200pt]{0.400pt}{394.353pt}}
\put(153,204){\usebox{\plotpoint}}
\put(153,203.67){\rule{3.132pt}{0.400pt}}
\multiput(153.00,203.17)(6.500,1.000){2}{\rule{1.566pt}{0.400pt}}
\put(180,204.67){\rule{3.132pt}{0.400pt}}
\multiput(180.00,204.17)(6.500,1.000){2}{\rule{1.566pt}{0.400pt}}
\put(166.0,205.0){\rule[-0.200pt]{3.373pt}{0.400pt}}
\put(206,205.67){\rule{3.132pt}{0.400pt}}
\multiput(206.00,205.17)(6.500,1.000){2}{\rule{1.566pt}{0.400pt}}
\put(193.0,206.0){\rule[-0.200pt]{3.132pt}{0.400pt}}
\put(232,206.67){\rule{3.373pt}{0.400pt}}
\multiput(232.00,206.17)(7.000,1.000){2}{\rule{1.686pt}{0.400pt}}
\put(219.0,207.0){\rule[-0.200pt]{3.132pt}{0.400pt}}
\put(272,207.67){\rule{3.132pt}{0.400pt}}
\multiput(272.00,207.17)(6.500,1.000){2}{\rule{1.566pt}{0.400pt}}
\put(246.0,208.0){\rule[-0.200pt]{6.263pt}{0.400pt}}
\put(325,208.67){\rule{3.132pt}{0.400pt}}
\multiput(325.00,208.17)(6.500,1.000){2}{\rule{1.566pt}{0.400pt}}
\put(285.0,209.0){\rule[-0.200pt]{9.636pt}{0.400pt}}
\put(364,209.67){\rule{3.373pt}{0.400pt}}
\multiput(364.00,209.17)(7.000,1.000){2}{\rule{1.686pt}{0.400pt}}
\put(338.0,210.0){\rule[-0.200pt]{6.263pt}{0.400pt}}
\put(391,210.67){\rule{3.132pt}{0.400pt}}
\multiput(391.00,210.17)(6.500,1.000){2}{\rule{1.566pt}{0.400pt}}
\put(378.0,211.0){\rule[-0.200pt]{3.132pt}{0.400pt}}
\put(430,211.67){\rule{3.373pt}{0.400pt}}
\multiput(430.00,211.17)(7.000,1.000){2}{\rule{1.686pt}{0.400pt}}
\put(404.0,212.0){\rule[-0.200pt]{6.263pt}{0.400pt}}
\put(457,212.67){\rule{3.132pt}{0.400pt}}
\multiput(457.00,212.17)(6.500,1.000){2}{\rule{1.566pt}{0.400pt}}
\put(444.0,213.0){\rule[-0.200pt]{3.132pt}{0.400pt}}
\put(483,213.67){\rule{3.132pt}{0.400pt}}
\multiput(483.00,213.17)(6.500,1.000){2}{\rule{1.566pt}{0.400pt}}
\put(496,214.67){\rule{3.373pt}{0.400pt}}
\multiput(496.00,214.17)(7.000,1.000){2}{\rule{1.686pt}{0.400pt}}
\put(510,216.17){\rule{2.700pt}{0.400pt}}
\multiput(510.00,215.17)(7.396,2.000){2}{\rule{1.350pt}{0.400pt}}
\multiput(523.00,218.59)(0.728,0.489){15}{\rule{0.678pt}{0.118pt}}
\multiput(523.00,217.17)(11.593,9.000){2}{\rule{0.339pt}{0.400pt}}
\multiput(536.00,227.61)(2.695,0.447){3}{\rule{1.833pt}{0.108pt}}
\multiput(536.00,226.17)(9.195,3.000){2}{\rule{0.917pt}{0.400pt}}
\put(549,229.67){\rule{3.132pt}{0.400pt}}
\multiput(549.00,229.17)(6.500,1.000){2}{\rule{1.566pt}{0.400pt}}
\put(562,230.67){\rule{3.373pt}{0.400pt}}
\multiput(562.00,230.17)(7.000,1.000){2}{\rule{1.686pt}{0.400pt}}
\put(576,231.67){\rule{3.132pt}{0.400pt}}
\multiput(576.00,231.17)(6.500,1.000){2}{\rule{1.566pt}{0.400pt}}
\put(589,232.67){\rule{3.132pt}{0.400pt}}
\multiput(589.00,232.17)(6.500,1.000){2}{\rule{1.566pt}{0.400pt}}
\put(602,233.67){\rule{3.132pt}{0.400pt}}
\multiput(602.00,233.17)(6.500,1.000){2}{\rule{1.566pt}{0.400pt}}
\put(470.0,214.0){\rule[-0.200pt]{3.132pt}{0.400pt}}
\put(642,234.67){\rule{3.132pt}{0.400pt}}
\multiput(642.00,234.17)(6.500,1.000){2}{\rule{1.566pt}{0.400pt}}
\put(615.0,235.0){\rule[-0.200pt]{6.504pt}{0.400pt}}
\put(906,234.67){\rule{3.132pt}{0.400pt}}
\multiput(906.00,235.17)(6.500,-1.000){2}{\rule{1.566pt}{0.400pt}}
\put(655.0,236.0){\rule[-0.200pt]{60.466pt}{0.400pt}}
\put(1051,233.67){\rule{3.132pt}{0.400pt}}
\multiput(1051.00,234.17)(6.500,-1.000){2}{\rule{1.566pt}{0.400pt}}
\put(919.0,235.0){\rule[-0.200pt]{31.799pt}{0.400pt}}
\put(1183,232.67){\rule{3.132pt}{0.400pt}}
\multiput(1183.00,233.17)(6.500,-1.000){2}{\rule{1.566pt}{0.400pt}}
\put(1064.0,234.0){\rule[-0.200pt]{28.667pt}{0.400pt}}
\put(1341,231.67){\rule{3.132pt}{0.400pt}}
\multiput(1341.00,232.17)(6.500,-1.000){2}{\rule{1.566pt}{0.400pt}}
\put(1196.0,233.0){\rule[-0.200pt]{34.930pt}{0.400pt}}
\put(1354.0,232.0){\rule[-0.200pt]{25.535pt}{0.400pt}}
\put(153,1296){\usebox{\plotpoint}}
\multiput(153.00,1296.59)(0.950,0.485){11}{\rule{0.843pt}{0.117pt}}
\multiput(153.00,1295.17)(11.251,7.000){2}{\rule{0.421pt}{0.400pt}}
\multiput(166.00,1303.60)(1.943,0.468){5}{\rule{1.500pt}{0.113pt}}
\multiput(166.00,1302.17)(10.887,4.000){2}{\rule{0.750pt}{0.400pt}}
\multiput(180.00,1307.60)(1.797,0.468){5}{\rule{1.400pt}{0.113pt}}
\multiput(180.00,1306.17)(10.094,4.000){2}{\rule{0.700pt}{0.400pt}}
\multiput(193.00,1311.61)(2.695,0.447){3}{\rule{1.833pt}{0.108pt}}
\multiput(193.00,1310.17)(9.195,3.000){2}{\rule{0.917pt}{0.400pt}}
\put(206,1314.17){\rule{2.700pt}{0.400pt}}
\multiput(206.00,1313.17)(7.396,2.000){2}{\rule{1.350pt}{0.400pt}}
\multiput(219.00,1316.61)(2.695,0.447){3}{\rule{1.833pt}{0.108pt}}
\multiput(219.00,1315.17)(9.195,3.000){2}{\rule{0.917pt}{0.400pt}}
\put(232,1319.17){\rule{2.900pt}{0.400pt}}
\multiput(232.00,1318.17)(7.981,2.000){2}{\rule{1.450pt}{0.400pt}}
\put(246,1321.17){\rule{2.700pt}{0.400pt}}
\multiput(246.00,1320.17)(7.396,2.000){2}{\rule{1.350pt}{0.400pt}}
\put(259,1323.17){\rule{2.700pt}{0.400pt}}
\multiput(259.00,1322.17)(7.396,2.000){2}{\rule{1.350pt}{0.400pt}}
\put(272,1325.17){\rule{2.700pt}{0.400pt}}
\multiput(272.00,1324.17)(7.396,2.000){2}{\rule{1.350pt}{0.400pt}}
\put(285,1327.17){\rule{2.700pt}{0.400pt}}
\multiput(285.00,1326.17)(7.396,2.000){2}{\rule{1.350pt}{0.400pt}}
\put(298,1329.17){\rule{2.900pt}{0.400pt}}
\multiput(298.00,1328.17)(7.981,2.000){2}{\rule{1.450pt}{0.400pt}}
\put(312,1331.17){\rule{2.700pt}{0.400pt}}
\multiput(312.00,1330.17)(7.396,2.000){2}{\rule{1.350pt}{0.400pt}}
\put(325,1333.17){\rule{2.700pt}{0.400pt}}
\multiput(325.00,1332.17)(7.396,2.000){2}{\rule{1.350pt}{0.400pt}}
\put(338,1335.17){\rule{2.700pt}{0.400pt}}
\multiput(338.00,1334.17)(7.396,2.000){2}{\rule{1.350pt}{0.400pt}}
\put(351,1337.17){\rule{2.700pt}{0.400pt}}
\multiput(351.00,1336.17)(7.396,2.000){2}{\rule{1.350pt}{0.400pt}}
\put(364,1339.17){\rule{2.900pt}{0.400pt}}
\multiput(364.00,1338.17)(7.981,2.000){2}{\rule{1.450pt}{0.400pt}}
\put(378,1341.17){\rule{2.700pt}{0.400pt}}
\multiput(378.00,1340.17)(7.396,2.000){2}{\rule{1.350pt}{0.400pt}}
\put(391,1343.17){\rule{2.700pt}{0.400pt}}
\multiput(391.00,1342.17)(7.396,2.000){2}{\rule{1.350pt}{0.400pt}}
\multiput(404.00,1345.61)(2.695,0.447){3}{\rule{1.833pt}{0.108pt}}
\multiput(404.00,1344.17)(9.195,3.000){2}{\rule{0.917pt}{0.400pt}}
\put(417,1348.17){\rule{2.700pt}{0.400pt}}
\multiput(417.00,1347.17)(7.396,2.000){2}{\rule{1.350pt}{0.400pt}}
\multiput(430.00,1350.61)(2.918,0.447){3}{\rule{1.967pt}{0.108pt}}
\multiput(430.00,1349.17)(9.918,3.000){2}{\rule{0.983pt}{0.400pt}}
\multiput(444.00,1353.61)(2.695,0.447){3}{\rule{1.833pt}{0.108pt}}
\multiput(444.00,1352.17)(9.195,3.000){2}{\rule{0.917pt}{0.400pt}}
\multiput(457.00,1356.60)(1.797,0.468){5}{\rule{1.400pt}{0.113pt}}
\multiput(457.00,1355.17)(10.094,4.000){2}{\rule{0.700pt}{0.400pt}}
\multiput(470.00,1360.60)(1.797,0.468){5}{\rule{1.400pt}{0.113pt}}
\multiput(470.00,1359.17)(10.094,4.000){2}{\rule{0.700pt}{0.400pt}}
\multiput(483.00,1364.59)(1.378,0.477){7}{\rule{1.140pt}{0.115pt}}
\multiput(483.00,1363.17)(10.634,5.000){2}{\rule{0.570pt}{0.400pt}}
\multiput(496.00,1369.59)(1.026,0.485){11}{\rule{0.900pt}{0.117pt}}
\multiput(496.00,1368.17)(12.132,7.000){2}{\rule{0.450pt}{0.400pt}}
\multiput(510.58,1376.00)(0.493,0.536){23}{\rule{0.119pt}{0.531pt}}
\multiput(509.17,1376.00)(13.000,12.898){2}{\rule{0.400pt}{0.265pt}}
\multiput(523.58,1390.00)(0.493,2.519){23}{\rule{0.119pt}{2.069pt}}
\multiput(522.17,1390.00)(13.000,59.705){2}{\rule{0.400pt}{1.035pt}}
\multiput(536.58,1454.00)(0.493,0.814){23}{\rule{0.119pt}{0.746pt}}
\multiput(535.17,1454.00)(13.000,19.451){2}{\rule{0.400pt}{0.373pt}}
\multiput(549.58,1475.00)(0.493,0.536){23}{\rule{0.119pt}{0.531pt}}
\multiput(548.17,1475.00)(13.000,12.898){2}{\rule{0.400pt}{0.265pt}}
\multiput(562.00,1489.58)(0.704,0.491){17}{\rule{0.660pt}{0.118pt}}
\multiput(562.00,1488.17)(12.630,10.000){2}{\rule{0.330pt}{0.400pt}}
\multiput(576.00,1499.59)(0.950,0.485){11}{\rule{0.843pt}{0.117pt}}
\multiput(576.00,1498.17)(11.251,7.000){2}{\rule{0.421pt}{0.400pt}}
\multiput(589.00,1506.59)(1.378,0.477){7}{\rule{1.140pt}{0.115pt}}
\multiput(589.00,1505.17)(10.634,5.000){2}{\rule{0.570pt}{0.400pt}}
\multiput(602.00,1511.60)(1.797,0.468){5}{\rule{1.400pt}{0.113pt}}
\multiput(602.00,1510.17)(10.094,4.000){2}{\rule{0.700pt}{0.400pt}}
\multiput(615.00,1515.60)(1.797,0.468){5}{\rule{1.400pt}{0.113pt}}
\multiput(615.00,1514.17)(10.094,4.000){2}{\rule{0.700pt}{0.400pt}}
\multiput(628.00,1519.61)(2.918,0.447){3}{\rule{1.967pt}{0.108pt}}
\multiput(628.00,1518.17)(9.918,3.000){2}{\rule{0.983pt}{0.400pt}}
\put(642,1522.17){\rule{2.700pt}{0.400pt}}
\multiput(642.00,1521.17)(7.396,2.000){2}{\rule{1.350pt}{0.400pt}}
\put(655,1523.67){\rule{3.132pt}{0.400pt}}
\multiput(655.00,1523.17)(6.500,1.000){2}{\rule{1.566pt}{0.400pt}}
\put(668,1525.17){\rule{2.700pt}{0.400pt}}
\multiput(668.00,1524.17)(7.396,2.000){2}{\rule{1.350pt}{0.400pt}}
\put(681,1526.67){\rule{3.132pt}{0.400pt}}
\multiput(681.00,1526.17)(6.500,1.000){2}{\rule{1.566pt}{0.400pt}}
\put(708,1527.67){\rule{3.132pt}{0.400pt}}
\multiput(708.00,1527.17)(6.500,1.000){2}{\rule{1.566pt}{0.400pt}}
\put(694.0,1528.0){\rule[-0.200pt]{3.373pt}{0.400pt}}
\put(734,1528.67){\rule{3.132pt}{0.400pt}}
\multiput(734.00,1528.17)(6.500,1.000){2}{\rule{1.566pt}{0.400pt}}
\put(721.0,1529.0){\rule[-0.200pt]{3.132pt}{0.400pt}}
\put(774,1528.67){\rule{3.132pt}{0.400pt}}
\multiput(774.00,1529.17)(6.500,-1.000){2}{\rule{1.566pt}{0.400pt}}
\put(747.0,1530.0){\rule[-0.200pt]{6.504pt}{0.400pt}}
\put(826,1527.67){\rule{3.373pt}{0.400pt}}
\multiput(826.00,1528.17)(7.000,-1.000){2}{\rule{1.686pt}{0.400pt}}
\put(787.0,1529.0){\rule[-0.200pt]{9.395pt}{0.400pt}}
\put(853,1526.67){\rule{3.132pt}{0.400pt}}
\multiput(853.00,1527.17)(6.500,-1.000){2}{\rule{1.566pt}{0.400pt}}
\put(840.0,1528.0){\rule[-0.200pt]{3.132pt}{0.400pt}}
\put(879,1525.67){\rule{3.132pt}{0.400pt}}
\multiput(879.00,1526.17)(6.500,-1.000){2}{\rule{1.566pt}{0.400pt}}
\put(866.0,1527.0){\rule[-0.200pt]{3.132pt}{0.400pt}}
\put(906,1524.67){\rule{3.132pt}{0.400pt}}
\multiput(906.00,1525.17)(6.500,-1.000){2}{\rule{1.566pt}{0.400pt}}
\put(892.0,1526.0){\rule[-0.200pt]{3.373pt}{0.400pt}}
\put(932,1523.67){\rule{3.132pt}{0.400pt}}
\multiput(932.00,1524.17)(6.500,-1.000){2}{\rule{1.566pt}{0.400pt}}
\put(945,1522.67){\rule{3.132pt}{0.400pt}}
\multiput(945.00,1523.17)(6.500,-1.000){2}{\rule{1.566pt}{0.400pt}}
\put(919.0,1525.0){\rule[-0.200pt]{3.132pt}{0.400pt}}
\put(972,1521.67){\rule{3.132pt}{0.400pt}}
\multiput(972.00,1522.17)(6.500,-1.000){2}{\rule{1.566pt}{0.400pt}}
\put(958.0,1523.0){\rule[-0.200pt]{3.373pt}{0.400pt}}
\put(998,1520.67){\rule{3.132pt}{0.400pt}}
\multiput(998.00,1521.17)(6.500,-1.000){2}{\rule{1.566pt}{0.400pt}}
\put(1011,1519.67){\rule{3.132pt}{0.400pt}}
\multiput(1011.00,1520.17)(6.500,-1.000){2}{\rule{1.566pt}{0.400pt}}
\put(985.0,1522.0){\rule[-0.200pt]{3.132pt}{0.400pt}}
\put(1038,1518.67){\rule{3.132pt}{0.400pt}}
\multiput(1038.00,1519.17)(6.500,-1.000){2}{\rule{1.566pt}{0.400pt}}
\put(1024.0,1520.0){\rule[-0.200pt]{3.373pt}{0.400pt}}
\put(1064,1517.67){\rule{3.132pt}{0.400pt}}
\multiput(1064.00,1518.17)(6.500,-1.000){2}{\rule{1.566pt}{0.400pt}}
\put(1077,1516.67){\rule{3.132pt}{0.400pt}}
\multiput(1077.00,1517.17)(6.500,-1.000){2}{\rule{1.566pt}{0.400pt}}
\put(1051.0,1519.0){\rule[-0.200pt]{3.132pt}{0.400pt}}
\put(1104,1515.67){\rule{3.132pt}{0.400pt}}
\multiput(1104.00,1516.17)(6.500,-1.000){2}{\rule{1.566pt}{0.400pt}}
\put(1117,1514.67){\rule{3.132pt}{0.400pt}}
\multiput(1117.00,1515.17)(6.500,-1.000){2}{\rule{1.566pt}{0.400pt}}
\put(1090.0,1517.0){\rule[-0.200pt]{3.373pt}{0.400pt}}
\put(1143,1513.67){\rule{3.132pt}{0.400pt}}
\multiput(1143.00,1514.17)(6.500,-1.000){2}{\rule{1.566pt}{0.400pt}}
\put(1156,1512.67){\rule{3.373pt}{0.400pt}}
\multiput(1156.00,1513.17)(7.000,-1.000){2}{\rule{1.686pt}{0.400pt}}
\put(1130.0,1515.0){\rule[-0.200pt]{3.132pt}{0.400pt}}
\put(1183,1511.67){\rule{3.132pt}{0.400pt}}
\multiput(1183.00,1512.17)(6.500,-1.000){2}{\rule{1.566pt}{0.400pt}}
\put(1196,1510.67){\rule{3.132pt}{0.400pt}}
\multiput(1196.00,1511.17)(6.500,-1.000){2}{\rule{1.566pt}{0.400pt}}
\put(1170.0,1513.0){\rule[-0.200pt]{3.132pt}{0.400pt}}
\put(1222,1509.67){\rule{3.373pt}{0.400pt}}
\multiput(1222.00,1510.17)(7.000,-1.000){2}{\rule{1.686pt}{0.400pt}}
\put(1209.0,1511.0){\rule[-0.200pt]{3.132pt}{0.400pt}}
\put(1249,1508.67){\rule{3.132pt}{0.400pt}}
\multiput(1249.00,1509.17)(6.500,-1.000){2}{\rule{1.566pt}{0.400pt}}
\put(1236.0,1510.0){\rule[-0.200pt]{3.132pt}{0.400pt}}
\put(1275,1507.67){\rule{3.132pt}{0.400pt}}
\multiput(1275.00,1508.17)(6.500,-1.000){2}{\rule{1.566pt}{0.400pt}}
\put(1288,1506.67){\rule{3.373pt}{0.400pt}}
\multiput(1288.00,1507.17)(7.000,-1.000){2}{\rule{1.686pt}{0.400pt}}
\put(1262.0,1509.0){\rule[-0.200pt]{3.132pt}{0.400pt}}
\put(1315,1505.67){\rule{3.132pt}{0.400pt}}
\multiput(1315.00,1506.17)(6.500,-1.000){2}{\rule{1.566pt}{0.400pt}}
\put(1302.0,1507.0){\rule[-0.200pt]{3.132pt}{0.400pt}}
\put(1341,1504.67){\rule{3.132pt}{0.400pt}}
\multiput(1341.00,1505.17)(6.500,-1.000){2}{\rule{1.566pt}{0.400pt}}
\put(1328.0,1506.0){\rule[-0.200pt]{3.132pt}{0.400pt}}
\put(1368,1503.67){\rule{3.132pt}{0.400pt}}
\multiput(1368.00,1504.17)(6.500,-1.000){2}{\rule{1.566pt}{0.400pt}}
\put(1381,1502.67){\rule{3.132pt}{0.400pt}}
\multiput(1381.00,1503.17)(6.500,-1.000){2}{\rule{1.566pt}{0.400pt}}
\put(1354.0,1505.0){\rule[-0.200pt]{3.373pt}{0.400pt}}
\put(1420,1501.67){\rule{3.373pt}{0.400pt}}
\multiput(1420.00,1502.17)(7.000,-1.000){2}{\rule{1.686pt}{0.400pt}}
\put(1434,1500.67){\rule{3.132pt}{0.400pt}}
\multiput(1434.00,1501.17)(6.500,-1.000){2}{\rule{1.566pt}{0.400pt}}
\put(1394.0,1503.0){\rule[-0.200pt]{6.263pt}{0.400pt}}
\put(1447.0,1501.0){\rule[-0.200pt]{3.132pt}{0.400pt}}
\put(153,842){\usebox{\plotpoint}}
\put(153,841.67){\rule{3.132pt}{0.400pt}}
\multiput(153.00,841.17)(6.500,1.000){2}{\rule{1.566pt}{0.400pt}}
\put(206,842.67){\rule{3.132pt}{0.400pt}}
\multiput(206.00,842.17)(6.500,1.000){2}{\rule{1.566pt}{0.400pt}}
\put(166.0,843.0){\rule[-0.200pt]{9.636pt}{0.400pt}}
\put(325,843.67){\rule{3.132pt}{0.400pt}}
\multiput(325.00,843.17)(6.500,1.000){2}{\rule{1.566pt}{0.400pt}}
\put(219.0,844.0){\rule[-0.200pt]{25.535pt}{0.400pt}}
\put(430,844.67){\rule{3.373pt}{0.400pt}}
\multiput(430.00,844.17)(7.000,1.000){2}{\rule{1.686pt}{0.400pt}}
\put(338.0,845.0){\rule[-0.200pt]{22.163pt}{0.400pt}}
\put(496,845.67){\rule{3.373pt}{0.400pt}}
\multiput(496.00,845.17)(7.000,1.000){2}{\rule{1.686pt}{0.400pt}}
\put(444.0,846.0){\rule[-0.200pt]{12.527pt}{0.400pt}}
\put(523,846.67){\rule{3.132pt}{0.400pt}}
\multiput(523.00,846.17)(6.500,1.000){2}{\rule{1.566pt}{0.400pt}}
\put(536,846.67){\rule{3.132pt}{0.400pt}}
\multiput(536.00,847.17)(6.500,-1.000){2}{\rule{1.566pt}{0.400pt}}
\put(510.0,847.0){\rule[-0.200pt]{3.132pt}{0.400pt}}
\put(615,845.67){\rule{3.132pt}{0.400pt}}
\multiput(615.00,846.17)(6.500,-1.000){2}{\rule{1.566pt}{0.400pt}}
\put(549.0,847.0){\rule[-0.200pt]{15.899pt}{0.400pt}}
\put(721,844.67){\rule{3.132pt}{0.400pt}}
\multiput(721.00,845.17)(6.500,-1.000){2}{\rule{1.566pt}{0.400pt}}
\put(628.0,846.0){\rule[-0.200pt]{22.404pt}{0.400pt}}
\put(853,843.67){\rule{3.132pt}{0.400pt}}
\multiput(853.00,844.17)(6.500,-1.000){2}{\rule{1.566pt}{0.400pt}}
\put(734.0,845.0){\rule[-0.200pt]{28.667pt}{0.400pt}}
\put(1051,842.67){\rule{3.132pt}{0.400pt}}
\multiput(1051.00,843.17)(6.500,-1.000){2}{\rule{1.566pt}{0.400pt}}
\put(866.0,844.0){\rule[-0.200pt]{44.566pt}{0.400pt}}
\put(1341,841.67){\rule{3.132pt}{0.400pt}}
\multiput(1341.00,842.17)(6.500,-1.000){2}{\rule{1.566pt}{0.400pt}}
\put(1064.0,843.0){\rule[-0.200pt]{66.729pt}{0.400pt}}
\put(1354.0,842.0){\rule[-0.200pt]{25.535pt}{0.400pt}}
\end{picture}
\vspace{0.5cm}
\caption{Direct $CP$ asymmetries of $B\rightarrow D^{(*)}
\bar{D}^{(*)}$ in the factorization approximation.}
\end{figure}
\begin{enumerate}
\item	All $CP$ asymmetries have the same sign and undergo
a change of magnitude at $k^2 =4m^2_c \approx 0.3m^2_b$. 
The asymmetry ${\cal A} (D\bar{D})$ 
is most sensitive to the uncertain penguin 
momentum transfer $k^2$, but its magnitude increases
only about $0.3\%$ from $k^2=0.01m^2_b$ to $k^2=m^2_b$. It is found
that the strong (gluonic) penguin effect is dominant over
the electroweak penguin effect, thus the latter is safely
negligible.

\item 	The $CP$ asymmetry ${\cal A}(D\bar{D})$ can be as 
large as $3\%$, while ${\cal A}(D\bar{D}^*)$ is only about
$2 \times 10^{-3}$. The smallness of the latter comes from the
cancellation effect, induced by the factor $(1+\xi)$ with
$\xi \sim -0.8$, in $S_u$ and $S_c$.
In our factorization approximation, the
asymmetries ${\cal A}(D^*\bar{D})$ and
${\cal A}(D^*\bar{D}^*)$ are identical and of magnitude
$1\%$. 

\item	Observation of the $CP$ asymmetries ${\cal A}(D\bar{D})$
and ${\cal A}(D^*\bar{D}^*)$ to three standard deviations
needs about $10^8$ $B^{\pm}_u$ events, if the composite
detection efficiency is at the $10\%$ level. 
More events are required to measure the same $CP$ asymmetries in $B_d$ decays,  
due to the cost for flavor tagging. 
\end{enumerate}
It is therefore worth while to search for such direct $CP$-violating
signals in the first-round experiments of $B$-meson factories.

\section{Indirect $CP$ violation}

Although the final-state rescattering phases have no effect on
direct $CP$ asymmetries ${\cal A}$  in our factorization
scheme, they are possible to influence the indirect $CP$ violation arising
from the interference between direct $B_d$ transition and 
$B^0_d$-$\bar{B}^0_d$ mixing in the decay modes under consideration. 
The characteristic measurable 
of this source of $CP$ violation is in general a difference between 
two rephasing-invariant quantities defined as \cite{Du}
\begin{eqnarray}
\Delta (f) & = & {\rm Im} \left [\frac{q}{p} \cdot
\frac{A(\bar{B}^0_d\rightarrow f)}{A(B^0_d\rightarrow f)}
\right ] \; , \nonumber \\
\bar{\Delta} ({\bar f}) & = & {\rm Im} \left [\frac{p}{q}
\cdot \frac{A(B^0_d\rightarrow \bar{f})}
{A(\bar{B}^0_d\rightarrow \bar{f})} \right ] \; ,
\end{eqnarray}
where $q/p = (V^*_{tb}V_{td})/(V_{tb}V^*_{td})$ denotes the
weak phase of $B^0_d$-$\bar{B}^0_d$ mixing 
\footnote{The $CP$ violation induced solely by $B^0_d$-$\bar{B}^0_d$
mixing (i.e., $|q/p|\neq 1$) is expected to be negligibly small
(of order $10^{-3}$ or smaller \cite{XingZZ97}) in the standard model.},
and $\bar{f}$ is the $CP$-conjugate state of $f$. If $f$ is a
$CP$ eigenstate (i.e., $|\bar{f}\rangle = CP |f\rangle =
\pm |f\rangle$) and the decay is dominated by the tree-level
channel, then $\bar{\Delta}({\bar f}) = - \Delta (f)$ is a good
approximation. In general only the difference $\bar{\Delta} ({\bar f}) 
- \Delta (f)$, which will vanish if all the CKM factors are real,
measures the $CP$ asymmetry. Note that the $CP$-even and
$CP$-odd components of $f = D^{*+}D^{*-}$ state or $f=D^{*0} \bar{D}^{*0}$
state may cause some dilution in the measurables $\Delta (f)$
and $\bar{\Delta}(\bar{f})$.
A proper treatment of indirect $CP$ violation in such 
modes is to make use of the angular analysis \cite{Dunietz}.
Alternatively one may evaluate the $P$-wave contribution to
$\Delta (D^{*+}D^{*-})$ and $\Delta (D^{*0}\bar{D}^{*0})$ by use of
the factorization approximation and the heavy quark symmetry \cite{Xing99}.

\vspace{0.4cm}

As penguin contributions to the transition amplitudes of 
$B\rightarrow D^{(*)}\bar{D}^{(*)}$ decays have been estimated to
be at the percent level, we expect that their effects on
$\Delta (f)$ and $\bar{\Delta}(\bar{f})$ are unimportant
and negligible.

\vspace{0.4cm}

It is obvious, as shown in Eq. (7), that for $B^0_d$
and $\bar{B}^0_d$ decays into the $CP$ eigenstates 
$D^+D^-$ and $D^0\bar{D}^0$ the amplitude ratios $\bar{A}^{+-}/A^{+-}$
and $\bar{A}^{00}/A^{00}$ are independent of the rescattering effects. 
Neglecting small penguin contributions
to $S_u$ and $S_c$ (i.e., taking $S_u =0$ and
$S_c =\bar{c}_2 + \bar{c}_1/3$), we arrive at
\begin{eqnarray}
\Delta (D^+D^-) & = & \Delta (D^0\bar{D}^0) \; =\;
+ \sin 2\beta \; , \nonumber \\
\bar{\Delta} (D^+D^-) & = & \bar{\Delta} (D^0\bar{D}^0)
\; =\; - \sin 2\beta \; ,
\end{eqnarray}
where $\beta$ is just the inner angle of the unitarity triangle 
defined in Eq. (1). Therefore the measurement of indirect 
$CP$ asymmetries in $B_d \rightarrow D^+D^-$ and
$B_d\rightarrow D^0\bar{D}^0$ decays may serve as a cross-check of 
$\sin 2\beta$ extracted from the $CP$ asymmetry in $B_d \rightarrow J/\psi K_S$
decays.

\vspace{0.4cm}

The channels $B_d \rightarrow D^{*+}D^-$, $D^+D^{*-}$ and
$B_d \rightarrow D^{*0}\bar{D}^0$, $D^0\bar{D}^{*0}$, whose
final states are non-$CP$ eigenstates, are also useful for
extraction of the weak angle $\beta$. Since the pseudoscalar and
vector mesons from $B^0_d$ and those from $\bar{B}_d$ have 
different quark-diagram configurations, the hadronic matrix elements 
and final-state rescattering phases in these two processes should 
in general be different \cite{Xing98}. As a result, 
\begin{eqnarray}
\Delta (D^{*+}D^-) & = & \zeta R_{+-} \sin (\delta + 2\beta)
\; , \nonumber \\
\bar{\Delta}(D^+D^{*-}) & = & \zeta R_{+-} \sin (\delta - 2\beta)
\; ;
\end{eqnarray}
and
\begin{eqnarray}
\Delta (D^{*0}\bar{D}^0) & = & \zeta R_{00} \sin (\delta + 2\beta)
\; , \nonumber \\
\bar{\Delta}(D^0\bar{D}^{*0}) & = & \zeta R_{00} \sin (\delta - 2\beta)
\; ,
\end{eqnarray}
where
\begin{eqnarray}
\zeta & = & \frac{Z_{D\bar{D}^*}}{Z_{D^*\bar{D}}} \; =\;
\frac{f_D}{f_{D^*}} \cdot \frac{A^{BD^*}_0 (m^2_D)}
{F^{BD}_1 (m^2_{D^*})} \; \;, 
\nonumber \\
\delta & = & \frac{\delta^{D\bar{D}^*}_1 + \delta^{D\bar{D}^*}_0}{2}
- \frac{\delta^{D^*\bar{D}}_1 + \delta^{D^*\bar{D}}_0}{2}
\; ;
\end{eqnarray}
and
\begin{eqnarray}
R_{+-} & = & \frac{\displaystyle 
\cos \frac {\delta^{D\bar{D}^*}_1 - \delta^{D\bar{D}^*}_0}{2}}
{\displaystyle \cos \frac{\delta^{D^*\bar{D}}_1 - \delta^{D^*\bar{D}}_0}{2}}
\; \; , 
\nonumber \\ \nonumber \\
R_{00} & = & \frac{\displaystyle 
\sin \frac{\delta^{D\bar{D}^*}_1 - \delta^{D\bar{D}^*}_0}
{2}}{\displaystyle \sin \frac{\delta^{D^*\bar{D}}_1 - \delta^{D^*\bar{D}}_0}{2}}
\; \; .
\end{eqnarray}
In obtaining these results we have neglected the small penguin
effects. The decay constants and form-factors in the expression of $\zeta$,
coming from decomposition of the hadronic matrix elements
$Z_{D\bar{D}^*}$ and $Z_{D^*\bar{D}}$ given in Eq. (5), are
self-explanatory. Note that $R_{+-} =1$ and $\delta = 
\delta^{D\bar{D}^*}_1 - \delta^{D^*\bar{D}}_1$ hold, if
one takes the limit $\delta^f_1 = \delta^f_0$ (for each 
final state $f$), in which the decay modes $B_d \rightarrow
D^{(*)0}\bar{D}^{(*)0}$ become forbidden. In the presence
of rescattering effects, i.e., $R_{+-}\neq 1$,
the extraction of $\beta$ from
$\Delta (D^{*+}D^-)$ and $\bar{\Delta}(D^+D^{*-})$ seems 
difficult. However, it is possible to
determine the isospin phase difference
$\delta^f_1 - \delta^f_0$ from the triangle relation in
Eq. (4), if the relevant rates of three 
(one charged $B$ and two neutral $B$) decay modes
are measured in experiments. The observation of
$B_d \rightarrow D^{*0}\bar{D}^0$ and $D^0\bar{D}^{*0}$
transitions turns out to be crucial: (a) if their 
branching ratios in comparison with
those of $B_d \rightarrow D^{*+}D^-$ and $D^+D^{*-}$ decays are too small to
be detected, then the final-state rescattering
effects should be negligible and the naive factorization
approach with $R_{+-}=1$ might work well; (b) if their branching
ratios are more or less comparable with those of
$B_d \rightarrow D^{*+}D^-$ and $D^+D^{*-}$ decays, then a quantitative
isospin analysis should be available, allowing 
us to extract the isospin phase differences and
determine the magnitudes of $R_{+-}$ and $R_{00}$.
In both cases, $\zeta$ can experimentally be determined 
and the result can be confronted with the theoretical
value of $\zeta$ calculated by inputting relevant
decay constants and form-factors.

\vspace{0.4cm}

For $B_d\rightarrow D^{*+}D^{*-}$ and $B_d\rightarrow D^{*0}
\bar{D}^{*0}$ decay modes the indirect $CP$ asymmetries need
a more careful analysis. Note that the transition amplitude
of $B^0_d\rightarrow D^{*+}D^{*-}$ (or $D^{*0}\bar{D}^{*0}$)
is a sum of three different components, i.e., 
the $S$-, $D$- and $P$-wave amplitudes \cite{Valencia}.
Without loss of generality the hadronic matrix elements $Z$
and $\bar{Z}$ for $B_d\rightarrow D^{*+}D^{*-}$ 
can be written as
\begin{eqnarray}
Z & = & \tilde{a} ~ (\epsilon_+ \cdot \epsilon_-) 
~ + ~ \frac{\tilde b}{m^2_{D^*}}
~ (p^{~}_0 \cdot \epsilon_+) (p^{~}_0 \cdot \epsilon_-)
\nonumber \\
&  & + ~ {\rm i} \frac{\tilde c}{m^2_{D^*}} ~ (\epsilon^{\alpha\beta\gamma\delta}
\epsilon_{+\alpha} \epsilon_{-\beta} p^{~}_{+\gamma}
p^{~}_{0\delta}) \; , \nonumber \\
\bar{Z} & = & {\tilde a} ~ (\epsilon_+ \cdot \epsilon_-) 
~ + ~ \frac{\tilde b}{m^2_{D^*}}
~ (p^{~}_0 \cdot \epsilon_+) (p^{~}_0 \cdot \epsilon_-)
\nonumber \\
&  & - ~ {\rm i} \frac{\tilde c}{m^2_{D^*}} ~ (\epsilon^{\alpha\beta\gamma\delta}
\epsilon_{+\alpha} \epsilon_{-\beta} p^{~}_{+\gamma}
p^{~}_{0\delta}) \; , 
\end{eqnarray}
where $\epsilon_{\pm}$ denotes the polarization of $D^{*\pm}$ meson,
$p^{~}_0$ and $p^{~}_{\pm}$ stand respectively for the momenta of $B_d$ and
$D^{*\pm}$ mesons, and $(\tilde{a},\tilde{b},\tilde{c})$ are
real scalars without the penguin effects. 
In terms of the decay constants and form factors,
$\tilde{a}$, $\tilde{b}$ and $\tilde{c}$ read explicitly as
\begin{eqnarray}
\tilde{a} & = & m^{~}_{D^*} f_{D^*} (m^{~}_B + m^{~}_{D^*})
A^{BD^*}_1(m^2_{D^*}) \; ,
\nonumber \\ \nonumber \\
\tilde{b} & = & -2 m^3_{D^*} f_{D^*} ~ \frac{A^{BD^*}_2(m^2_{D^*})}
{m^{~}_B + m^{~}_{D^*}} \; ,
\nonumber \\ \nonumber \\
\tilde{c} & = & -2 m^3_{D^*} f_{D^*} ~ \frac{V^{BD^*}(m^2_{D^*})}
{m^{~}_B + m^{~}_{D^*}} \; .
\end{eqnarray}
In the absence of angular analysis one may first calculate the
ratio $\bar{Z}/Z$ by summing over the polarizations of two
final-state vector mesons \cite{Valencia}, and then calculate the
$CP$-violating quantities $\Delta (D^{*+}D^{*-})$ and
$\bar{\Delta} (D^{*+}\bar{D}^{*-})$ in the neglect of small
penguin effects. 
We finally arrive at
\begin{eqnarray}
\Delta (D^{*+}D^{*-}) & = & \Delta (D^{*0}\bar{D}^{*0})
\; =\; + \sin 2\beta ~ \frac{ 1 - \chi}{1+\chi} \; ,
\nonumber \\
\bar{\Delta} (D^{*+}D^{*-}) & = & \Delta (D^{*0}\bar{D}^{*0})
\; =\; - \sin 2\beta ~ \frac{1-\chi}{1+\chi} \; ,
\end{eqnarray}
where 
\begin{equation}
\chi \; =\; \frac{2 (x^2 -1) \tilde{c}^2}
{(2+x^2)\tilde{a}^2 + (x^2-1)^2 \tilde{b}^2 + 2x (x^2-1) \tilde{a}
\tilde{b}} \; 
\end{equation}
with $x = (m^2_B - 2m^2_{D^*})/(2m^2_{D^*}) =2.45$. Clearly
the dilution parameter $\chi$ results from the $P$-wave
contribution to the overall decay amplitudes.
If we adopt the simple monopole model for relevant 
form factors \cite{Stech}, 
it turns out that $V^{BD^*}(m^2_{D^*}) =0.784$, $A^{BD^*}_1(m^2_{D^*})
=0.715$ and $A^{BD^*}_2(m^2_{D^*})=0.753$. Accordingly
$\tilde{b}/\tilde{a}= -0.160$ and $\tilde{c}/\tilde{a} =-0.167$.
The relationship $\tilde{b}/\tilde{a} \approx \tilde{c}/\tilde{a}$
is indeed guaranteed by the heavy quark symmetry,
which makes the form factors appearing in Eq. (23) related to one another.
In this symmetry limit we obtain \cite{Xing99} 
\begin{equation}
\frac{\tilde{b}}{\tilde{a}} \; =\;
\frac{\tilde{c}}{\tilde{a}} \; =\;
-\frac{2m^2_{D^*}}{m^{~}_B (m^{~}_B + 2m^{~}_{D^*})} \;\; ,
\end{equation}
amounting to $-0.164$. Then we get
$(1-\chi)/(1+\chi) \approx 0.89$, a value deviating only about 
$11\%$ from unity. 
Note that this dilution factor can also be determined
from measuring the ratio $\Delta (D^{*+}D^{*-}) / \Delta (D^+D^-)$.
From this estimation we find that the $P$-wave dilution effect
is not very significant, therefore extracting the $CP$-violating
parameter $\sin 2\beta$ from $B_d\rightarrow D^{*+}D^{*-}$ decays
remains possible even if a delicate angular analysis is not made.

\section{Summary}

We have analyzed direct and indirect $CP$ asymmetries in 
$B^0_d$ vs $\bar{B}^0_d \rightarrow D^+D^-$, $D^{*+}D^-$, $D^+D^{*-}$ and
$D^{*+}D^{*-}$ decays. The isospin amplitudes of these
transitions are calculated with the help of the effective weak
Hamiltonian and the factorization approximation, and the long-distance
interactions at the hadron level are taken in to account by
introducing elastic rescattering phases for two isospin channels
of the final-state mesons. We have shown that in this 
factorization approach the direct $CP$ violation is irrelevant
to the final-state rescattering effects, i.e., it is governed
only by the short-distance penguin mechanism. The magnitude of
direct $CP$ violation is estimated to be $3\%$ in $B_d\rightarrow
D^+D^-$ decay modes. The same amount of $CP$ violation can
manifest itself in the charged $B_u$ decays into $D^+\bar{D}^0$
and $D^-D^0$ states, which are easier to be measured at $B$-meson
factories. We have demonstrated that the penguin effects on 
indirect $CP$ asymmetries in $B_d\rightarrow D^{(*)+}D^{(*)-}$ 
decays are insignificant and even negligible. While the 
long-distance rescattering has no effect on indirect $CP$ violation
in $B_d \rightarrow D^+D^-$ and $D^{*+}D^{*-}$ transitions, it
may affect that in $B_d\rightarrow D^{\pm}D^{*\mp}$ modes, whose 
final states are non-$CP$ eigenstates. We have calculated the
$P$-wave contribution to the indirect $CP$ asymmetry in
$B_d\rightarrow D^{*+}D^{*-}$ decays. The corresponding dilution  
effect is found to be insignificant, therefore observation
of large $CP$ violation 
remains under expectation even without the delicate
angular analysis. 

\vspace{0.4cm}

It is certainly necessary to test the validity of our factorization
hypothesis, on which most of the afore-mentioned results depend.
To do so a measurement of $B_d\rightarrow D^{(*)0}\bar{D}^{(*)0}$
transitions will be particularly helpful. 
On the one hand,
if the branching ratios of these decay modes are too small 
comparied with those of $B_d\rightarrow D^{(*)+}D^{(*)-}$
transitions, then the final-state rescattering effects should be
negligible and the naive factorization approximation might work
well. On the other hand, if the branching ratios of $B_d$ decays
into $D^{(*)0}\bar{D}^{(*)0}$ and $D^{(*)+}D^{(*)-}$ states are
found to be more or less comparable in magnitude, then a
quantitative isospin analysis should become available, allowing
us to extract the isospin phase differences and control the
final-state rescattering effects. In any case much can be learnt
about the factorization hypothesis and its applicability in $B$
decays into two heavy charmed mesons.

\vspace{0.4cm}

In conclusion, the observation of direct and indirect
$CP$ asymmetries in $B_d\rightarrow D^{(*)+}D^{(*)-}$ decays
is promising at $B$-meson factories. They are expected to
provide us some valuable information
about the weak phase $\beta$ as well as the penguin and rescattering 
effects in nonleptonic $B$ decays.

\end{document}